\documentclass[amsmath,prb,twocolumn,showpacs]{revtex4}

\usepackage{graphicx}
\usepackage{bm}

\begin{document}
\title{Full Counting Statistics of Charge Transfer in Coulomb Blockade Systems}
\author{D.~A.~Bagrets and Yu.~V.~Nazarov}
\address{Department of Applied Physics and Delft Institute of Microelectronics
and Submicrontechnology, \\
Delft University of Technology, Lorentzweg 1, 2628 CJ Delft, The Netherlands
}
\date{\today}

\begin{abstract}
Full counting statistics (FCS) of charge transfer in mesoscopic systems has recently
become a subject of significant interest, since it proves to reveal an important information 
about the system which can be hardly assessed by
other means. While the previous research mostly addressed the FCS of non-interacting
systems, the present paper deals with the FCS in the limit of strong interaction.
In this Coulomb blockade limit the electron dynamics is known to be governed by a 
master equation.
We develop a general scheme to evaluate the FCS in such case, this being the main result
of the work presented. We illustrate the scheme, by applying it 
to concrete systems. For generic case of a single resonant level we
establish the equivalence of scattering and master equation approach to FCS. 
Further we study 
a single Coulomb blockade island with two and three leads attached and compare the
FCS in this case with our recent results concerning an open dot either with two and three
terminals. We demonstrate that Coulomb interaction suppresses the relative 
probabilities of large current fluctuations. 
\end{abstract}

\maketitle

\begin{section}{Introduction}
  The current fluctuations in the various mesoscopic systems have been
the subject of both theoretical and experimental research in the last two decades.  
Traditionally, the attention was focused on the shot noise phenomenon.
The shot noise is the main fundamental source of current noise at low temperatures. 
In classical systems shot noise unambiguously related to the discreteness of the electron 
charge. In quantum system the shot noise can be used as unique tool to reveal
the information about the electron correlations and entanglement of different kind. 
The investigation of the quantum shot noise cross-correlations 
in the multi-terminal mesoscopic devices is the new trend in this field which  
has attracted much attention as well.
The most achievements in the study of the shot noise phenomena have been summarized 
in the recent review article~\cite{BlanterReview}.

   Alternative way to investigate  the current correlations in the mesoscopic systems
has proposed in the pioneering work by Levitov {\it et. al.}~\cite{Levitov}.
This new fascinating theoretical
approach, known as the {\it full counting statistics}, yields not only shot noise power
but also all possible correlations and momenta of charge transfer.
The essence of this method is an evaluation of the probability distribution function
of the numbers of electrons transferred to the given terminals during 
the given period of time. The first and the second moments of this distribution
correspond to the average currents and the shot-noise correlations, respectively.
The probability distribution also contains the fundamental information about 
large current fluctuations in the system.

  Initially, FCS method~\cite{Levitov} made use of the scattering approach to 
mesoscopic transport. It was assumed that the mesoscopic system was completely 
characterized by its scattering matrix.  This method enabled the study the
statistics of the transport through the disordered metallic conductor~\cite{Yakovets}
and the two-terminal chaotic cavity~\cite{Blanter}. Muzykantskii and Khmelnitskii
generalized the original approach to the case of the normal metal/superconducting
contacts. The very recent development in this field is the 
counting statistics of the charge pumping in the open quantum 
dots~\cite{Andreev, Levitov1, Mirlin}.

   The use of multichannel scattering matrix of the system was crucial to obtain the
results of the above mentioned works. However, such approach leads to
the difficulties in case of practical layouts, where the scattering matrix is random and
cumbersome. They become apparent especially in case of multi-terminal geometry.
To circumvent these difficulties one evaluates the FCS with the semiclassical Keldysh Green
function method~\cite{RefYuli} or with its simplification called  the circuit theory of mesoscopic 
transport~\cite{General}. The Keldysh method to FCS was first proposed by one of the authors in
order to treat the effects of the weak localization corrections onto the FCS in the disordered
metallic wires. The method proves to be very flexible and has been recently applied 
to the FCS in superconducting heterostructures~\cite{Belzig}, multi-terminal normal metal
systems~\cite{NazBag} as well as in the three-terminal superconducting beam splitter~\cite{Borlin}.

  The above research addressed the FCS of non-interacting electrons. Since the interaction
may  bring correlations and entanglement of electron states the study of FCS of
interacting electrons is both challenging and interesting. 
In this paper we present an extensive theory of FCS in mesoscopic systems placed in 
a strong Coulomb blockade limit. 
  
   Note,  that the shot-noise in the Coulomb blockade devices has 
attracted the significant attention. Korotkov~\cite{Korotkov} and Hershfield 
{\it et.al.}~\cite{Hershfield} presented the first theory in the framework of
"orthodox" approach to single electron transport.
Later on Korotkov also studied the frequency dependence 
of the shot noise by means of Langevin approach both in low (classical) and 
very high (quantum) frequency limits. The frequency dependence of the shot noise
in the single electron transistor was also investigated in Ref.~\cite{Galperin, Hanke}
The ferromagnetic single electron transistor was considered by Bulka {\it et. al. }~\cite{Bulka}. 
The shot noise experiments were performed by Birk {\it et. al.}~\cite{Birk}. 
In this work, the nano-particle 
in between the STM (scanning tunneling microscope) tip and the metallic electrode was used to
form the Coulomb blockade island and the quantitative agreement with the theory of
Hershfield {\it et. al.} was found.

   The electrons dynamics in Coulomb blockade limit is fortunately relatively simple.  
When the cotunneling phenomena is disregarded, the evolution of the system is governed
by a master equation. The charge transfer is thus a classical stochastic
process rather than the quantum mechanical one. Nevertheless the FCS is by no means trivial
and has not been studied yet. In the present paper we have developed the general approach to FCS 
in the Coulomb blockade regime. This is the central result of the paper. Our method turns out to 
be an elegant extension of the usual master equation approach. 
We also predict the FCS in different Coulomb blockade systems. The previous results for the 
zero frequency shot noise power can be evaluated in our approach as the second moment of 
charge transfer probability distribution function.

   The paper is organized as follows. In the section II we start by presenting
the two physical systems to be treated within the master equation.  
Basing on these prototypes we formulate the general model. We derive our approach 
to the FCS in the section III.  In the section IV we applied to the single resonant level 
and consider its relation to the scattering approach. Two- and three-terminal single
electron transistors are considered in section V. 
We also compare their FCS in the Coulomb blockade limit with that one of non-interacting
electrons~\cite{NazBag}. We summarize the results in section VI.

\end{section}

\begin{section}{Systems under consideration}
The dynamics of many different physical systems can  be described by master equation. 
For our  purposes it is convenient to write it down as follows
\begin{equation}
 \frac{\partial}{\partial t}|p,t\rangle = -\hat{L} |p(t)\rangle
 \label{MasterEq}
\end{equation}
where each element $p_n(t)$ of the vector $|p(t)\rangle$ corresponds to the probability to
find the system in the  state $n$. The matrix elements of operator $\hat L$ are given by 
\begin{equation}
 L_{mn} = \delta_{nm}\gamma_n - \Gamma_{m\leftarrow n}, \quad
 \gamma_n = \sum_{m\ne n} \Gamma_{m\leftarrow n} \label{Lmatrix}
\end{equation}
Here $\Gamma_{n\leftarrow m}$  stands for the transition rate from the state $m$
to the state $n$.  $\gamma_n$ stands for the total transition rate from the state $n$. 
Thus defined operator $\hat L$ always has a zero eigenvalue, the corresponding 
eigenvector being the stationary solution of the master equation. 

A variety of Coulomb blockade mesoscopic systems obey Eq.~({\ref{MasterEq}}) 
under appropriate conditions. Here the main advantage of the master equation approach
is a possibility of
non-pertubative  treatment of the interaction effects. In what follows, we first remind 
the master equation description of two simple systems: single resonant level and 
many-terminal Coulomb 
blockade island. On the basis of these examples we will sketch the master equation for the 
general Coulomb blockade system. This will prepare us to the next section where we derive
the FCS method.

\begin{subsection} {Resonant level model.}
An elaborated model of the resonant center was presented in Ref.~\cite{LarMat}.
It was subsequently improved in the work~\cite{GlazMat} to include the Coulomb interaction.
One the physical realization is disordered tunneling barrier which is placed
between two leads.~\cite{Savchenko}. At sufficiently low
temperatures the main mechanism of transport in this system is the resonant tunneling
via localized states formed by impurity centers. Another physical realization is the Coulomb
blockade quantum dot in the low temperature regime $k_B T\ll \Delta E $,
where  $\Delta E$ is a mean separation between the
energy levels in the dot.~\cite{Beenakker} By applying the gate voltage,
one can tune the given level  to be between the chemical potentials of the leads.  
(See Fig.1). We consider below two limiting cases where one disregards either double occupancy
of the level or on-cite Coulomb interaction.  

\begin{figure}[t]
\includegraphics[scale=0.3]{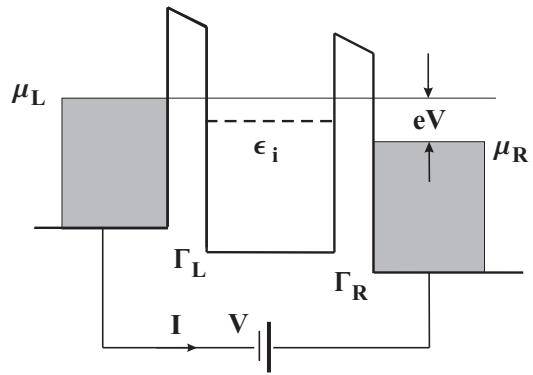}
\caption{ 
The single resonant level system, formed by the two tunnel barriers. The resonant
level in the quantum well is shown by the dashed line.}
\end{figure}

In the strongly interacting case the double occupancy of the resonant level is entirely 
excluded due to the Coulomb repulsion $U$. Then the system can be found only in two 
different microscopic states: one with no electrons, and another with a single electron.
The transport through the level can be described by master
equation approach, provided  the applied voltage or the temperature are not
too low, i.e. ${\rm max}\{eV,k_B T\}\gg \hbar \Gamma_{L(R)}$. Here $\Gamma_{L(R)}$
are the quantum-mechanical tunneling rates from the left (right) electrode onto
the resonant level. We will also assume that at the relevant energy scale, given
by ${\rm max}\{eV,k_B T\}$, the rates $\Gamma_{L(R)}$ are energy independent.
		
   Under above assumptions the transition rates in Eq. (1) are given by
\begin{eqnarray}
 \Gamma_{1\leftarrow 0} &=& 2\,\Gamma_L f_L(\epsilon_i) + 2\,\Gamma_R f_R(\epsilon_i) 
 \label{RatesU} \\
 \Gamma_{0\leftarrow 1} &=& \Gamma_L [1-f_L(\epsilon_i)] + \Gamma_R [1-f_R(\epsilon_i)] 
\nonumber
\end{eqnarray}
The microscopic states $\{0\}$ and $\{1\}$ denote the situation with no and one electron, 
respectively. Fermi function $f_{L(R)}(\epsilon) = (1+\exp[(\epsilon-\mu_{L(R)})/kT])^{-1}$
accounts for the filling in the left (right) lead and $\epsilon_i$ is the position of 
the resonant level. The factor 2 in the rate $\Gamma_{1\leftarrow 0}$ stems 
from the fact that two quantum states, with spin up and down, are available for tunneling.
The description in terms of rates~(\ref{RatesU}) is correct when the Coulomb repulsion
is strong enough: $U\gg{\rm max}\{eV,k_B T\}$.

   The opposite limit is the case of vanishing Coulomb interaction. In this case
the spin up and down channels can be treated independently. Each of them 
can be described by the master equation, provided
the same condition as before is fulfilled: ${\rm max}\{eV,k_B T\}\gg \hbar \Gamma_{L(R)}$.
For both spin directions the rates are written as
 \begin{eqnarray}
 \Gamma_{1\leftarrow 0} &=& \Gamma_L f_L(\epsilon_i) + \Gamma_R f_R(\epsilon_i) 
\label{Rates_non_int} \\
 \Gamma_{0\leftarrow 1} &=& \Gamma_L [1-f_L(\epsilon_i)] + \Gamma_R [1-f_R(\epsilon_i)] 
\nonumber
\end{eqnarray}
Here the indices $\{0\}$ and $\{1\}$ denote the filling factor of the level by electon with a 
chosen spin. 

  Disregarding of Coulomb interaction is not  
adequate for a realistic system. However the latter model is worth to consider as well.
The point is that the statistics of the charge transfer in this case can be also evaluated 
in the framework of the non-interacting scattering approach~\cite{Levitov}, thus providing
the way to establish the consistency of two approaches to FCS.
\end{subsection}

\begin{figure}[b]
\begin{center}
\includegraphics[scale=0.3]{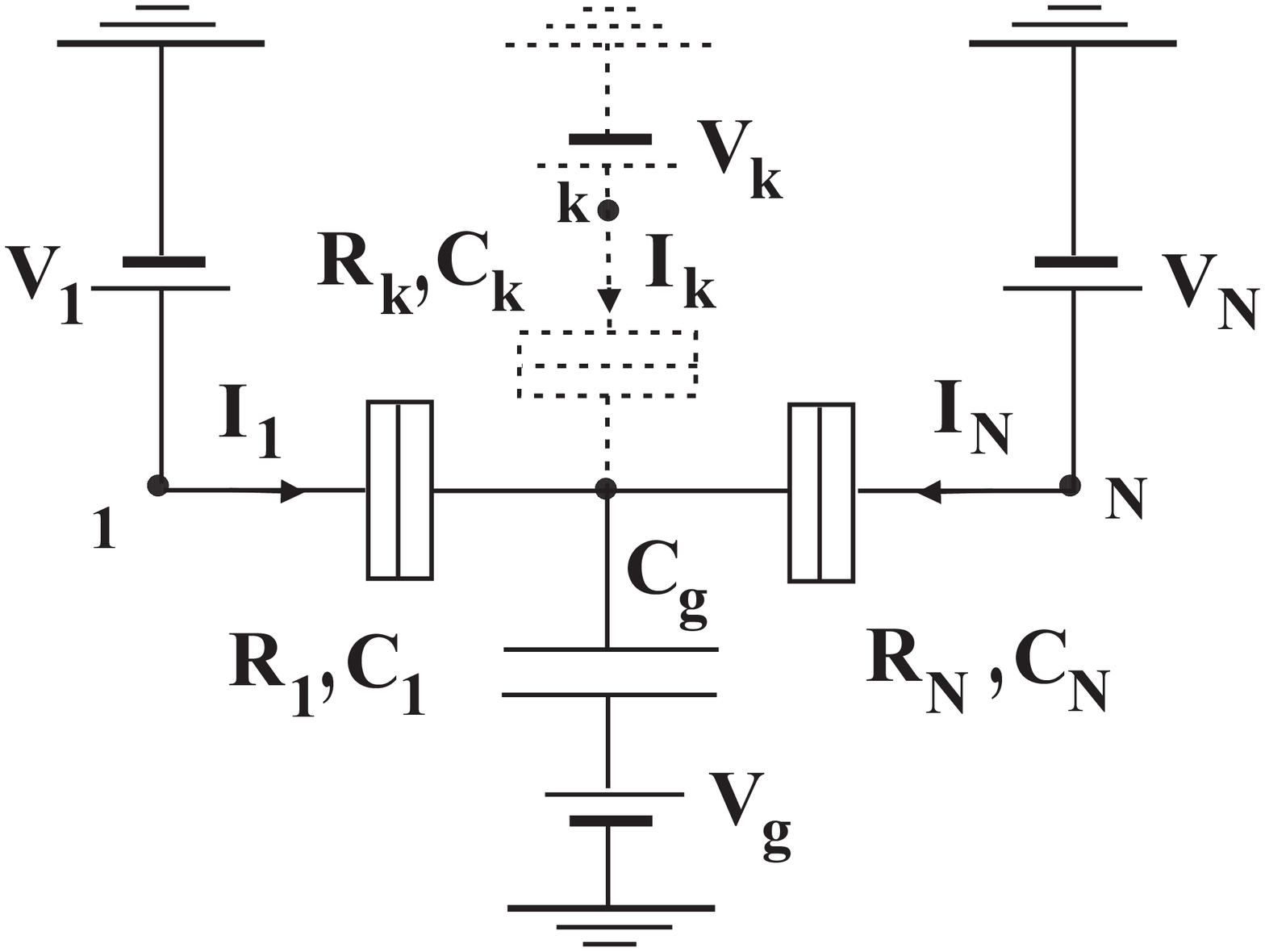}
\caption{The equivalent circuit of the N-terminal Coulomb blockade island. Each junction
$k$ is biased by the external voltage source $V_k$. The dot itself is capacitatively coupled
with the gate voltage $V_g$.} 
\end{center}
\end{figure}

\begin{subsection} {Many-terminal Coulomb blockade island.} 
The electrical circuit incorporating
the quantum dot with several terminals is shown in Fig.~2. This circuit is an
extension of the usual set-up in case of conventional two-terminal 
quantum dot.~\cite{IngoldNaz} 
At the present stage of nanotechnology the mesoscopic system, associated with this 
circuit, can be realized with the use of two-dimensional electron gas (2DEG) in the 
GaAs/AlGaAs heterostructures. 

  The essential elements of the circuit shown in Fig.2 are the resistances $R_k$
of the contacts, the mutual capacitances $C_k$ between the leads and the island and
the external dc voltage sources $V_k$. Correspondingly, $C_g$ and $V_g$ denote
the gate capacitance and gate voltage, which is used to vary the offset change
on the island. We assume that the dot is placed in the 
Coulomb blockade regime, $R_k \gg R_Q = 2\pi\hbar/{e^2}$.
In order the Coulomb blockade effect will be observable the condition 
$k_B T \ll E_c = e^2/2C_{\Sigma}$ is also required. Here $E_c$ is a charging
energy of the island, $C_{\Sigma} = \sum_{i=1}^N C_k + C_g$ is a sum capacitance of 
the system and $N\ge2$ is a number of leads attached to the dot.
We also assume the temperature to be rather high, 
$k_B T \gg \Delta E$, with $\Delta E$ being the mean level spacing in the dot, so that
the discreteness of the energy spectrum in the island is not important.
The possible effects of co-tunneling will not be discussed in the
paper. Therefore the characteristic scale of applied voltage $eV$ 
is assumed to be greater than the Coulomb blockade threshold, $eV\ge E_c$.

  Under the above conditions the multi-terminal dot is fairly well described
by the "orthodox" Coulomb blockade theory. One can consider the excess 
number of electrons on the island ($-n$) as a good quantum number, 
corresponding to the macroscopic state of the system. The tunneling of electrons will occur
one by one, increasing or decreasing the charge $Q_0=ne$ on the island by $\pm e$.
The corresponding tunneling rate $\Gamma_{n\pm 1\leftarrow n }^{(k)}$ across the junction $k$ 
is expressed via the electrostatic energy difference $\Delta E_{n\pm 1\leftarrow n }^{(k)}$
between the initial ($n$) and
final $(n\pm 1)$ configurations
\begin{equation}
 \Gamma_{n\pm 1\leftarrow n }^{(k)} = \frac{1}{e^2 R_k}
\frac{ \Delta E_{n\pm 1\leftarrow n }^{(k)} }
{1-\exp[-\Delta E_{n\pm 1\leftarrow n }^{(k)}/k_B T]}
\label{QRates}
\end{equation}
The evaluation of  $\Delta E_{n\pm 1\leftarrow n }^{(k)}$ can be done along the same lines
as in the case of two-terminal dot.~\cite{IngoldNaz} The result reads 
\begin{equation}
 \Delta E_{n\pm 1\leftarrow n }^{(k)} = \pm e \Bigl( V_k - V_0(n) \Bigr) - 
\frac{e^2}{2C_\Sigma}
\label{deltaE}
\end{equation}
where $V_0(n)$ is the electrostatic potential on the island. It is written as
\begin{equation}
 V_0(n) = \frac{1}{C_\Sigma}\Bigl(en+C_g V_g\Bigr) + 
\frac{1}{C_\Sigma}\sum_{i=1}^{N}{\widetilde C}_i V_i
\label{V0n}
\end{equation}
Here ${\widetilde C}_i \equiv C_i + C_g/N$ and we also assumed that the dot is biased
such a way, that external voltages are subjected to the condition $\sum_{i=1}^N V_i = 0$.
In this case the gate voltage $V_g$ can be used to influence the offset charge 
$q=C_g V_g$ on the island in a controlled way.

Neglecting the quantum correlations between different tunneling processes, we may 
write down the master equation~(\ref{MasterEq}). It connects the states with 
different island charge, the total transition rate from the state $n$ 
to $n\pm 1$ being the sum of tunneling rates over all junctions:
$\Gamma_{n\pm 1\leftarrow n } = \sum_{k=1}^N \Gamma_{n\pm 1\leftarrow n }^{(k)}$.

\end{subsection}

\begin{subsection}{Universal model.} 

\begin{figure}[b]
\begin{center}
\includegraphics[scale=0.22]{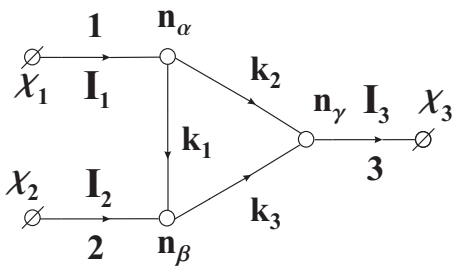}
\caption{The graph of universal model (See the main text). The terminals are connected
with the system via external junctions 1,2 and 3. The nodes $\alpha$, $\beta$ and $\gamma$
are either resonant levels or dots, linked with each other by internal junctions $k$'s.
The arrows denote the conventional direction of a current through each junction.
} 
\end{center}
\end{figure} 

 We now outline the universal model, which is an extension of
the preceding two. The physical realization of this model can be either
an array of Coulomb blockade quantum dots or a mesoscopic system with a number
of resonant levels. We assume that all the relevant conditions, mentioned previously,
are satisfied and therefore the description in terms of master equation is valid.
It is convenient schematically to represent the system as a graph (see Fig.~3),
so that each node $\alpha$ would correspond either to the single dot, the single resonant 
level or the external terminals 
and the line $(\alpha,\beta)$ would be associated with the possible transition.
Let $M$ be the total number of nodes in this graph.
We denote nodes by the Greek indices and numerate each line by the Latin 
indices, i.e. we will write $k=(\alpha,\beta)$, where $k=1 \dots L$ and $L$ is a total 
number of lines.  In case of many-dot system each line $k$ 
unambiguously corresponds to the tunnel junction.
In case of mesoscopic system with many resonant levels one has no 
a direct one-to-one correspondence between the line of the graph and the junction. 
Still, for the sake of convenience, we will sometimes refer below to each line of the graph 
as to the "junction" when it may not lead to contradiction. 
The lines are assumed to be directed, thus specifying 
the sign convention for a current $I_k$ through the "junction" $k$. There are $N$ external 
junction $k=1\dots N$, ($N\le L$), connecting the terminals with the system.  The currents 
through these junctions are directly measurable and hence are of our main 
interest in this paper.

  The macro- or microscopic state of the universal model is given by a set of occupation
numbers $|n\rangle = |n_1, \dots, n_M\rangle$;  $n_\alpha$ is equal to any
integer for the array of quantum dots and refers to the excess charge on the
island $\alpha$; in case of many resonant levels system $n_\alpha=0,1$, since
it is assumed that the double occupancy is totally suppressed due to the  strong 
Coulomb interaction.

  We now consider the general properties of the master equation~(\ref{MasterEq}) describing
the above model. Owing to the fact $\sum_n L_{nm} = 0$, the $\hat L$ operator has the right, 
$|p_0\rangle$, and the left, $\langle q_0|$, eigenvectors corresponding to zero eigenvalue
\begin{equation}
 \hat L |p_0\rangle = 0, \qquad \langle q_0 |\hat L = 0
\label{Equilib}
\end{equation}
We assume that they are unique, otherwise it would imply the situation
with two or more non-interacting subsystems.
The vector $|p_0\rangle$ gives the steady probability distribution and
$\langle q_0 | = (1, 1, \dots, 1)$. 
Since we are interested, in general, only in the permanent, but not the transient 
processes in the system it is naturally to restrict the consideration only
to absorbing states $n$. Thus we will exclude all the transient ones $n'$, for which
$\Gamma_{n n'}>0$ but at the same time $\Gamma_{n'n}=0$. We assume that the $\hat L$ operator,
bounded to the absorbing states, has a complete set of left and write eigenvectors
\begin{equation}
 \hat L |p_k \rangle = |p_k \rangle{\lambda}_k, 
 \quad {\lambda}_k \langle q_k | = \langle q_k |\hat L ,
 \quad \sum_{k} |p_k \rangle \langle q_k | = \hat I
\label{Eigen}
\end{equation}
where $\hat I$ is a unitary operator in the absorbing subspace.
For any physically reasonable $\hat L$ operator $\lambda_0 = 0$ and
${\rm Re}\,\lambda_k > 0$ for $k\ne 0$. 

For the following it is also useful to represent $\hat L$ operator in the form
\begin{equation}
\hat L = \hat \gamma - \hat \Gamma, \quad
\hat\Gamma = \sum_{k=1}^L (\hat\Gamma_k^{(+)} + \hat\Gamma_k^{(-)})
\end{equation}
where $\hat\gamma$ is the diagonal operator in the basis $|n\rangle$ of the system
configuration
and $\hat\Gamma_k^{(\pm)}$ are associated with the tunneling transitions through the
"junction" $k=(\alpha,\beta)$:
\begin{equation}
 \hat\gamma = \sum_{\{n\}} |n\rangle\gamma(n) \langle n|, \quad
  \hat\Gamma_k^{(\pm)} = \sum_{\{n\}} |n'\rangle\Gamma_k^{(\pm)}(n) \langle n|
\label{GOperators}
\end{equation}
 The state $|n'\rangle = |n_1,\dots,n'_\alpha,\dots, n'_\beta,\dots, n_M \rangle$ results
from the state $|n\rangle$  by appropriate changing the corresponding occupation numbers:
$n'_\alpha = n_\alpha - \sigma_k$,  $n'_\beta = n_\beta + \sigma_k$, where $\sigma_k=\pm 1$
denotes the direction of the transition. 
   
\end{subsection}

\end{section}

\begin{section}{The FCS in the master equation}

   In this section we derive the general scheme to evaluate the FCS  in the arbitrary
mesoscopic system placed in the strong Coulomb blockade regime. In what follows, 
we assume that the mesoscopic system under study can be described
by the universal model outlined in the preceding section.  We also assume that
the transition rates of the corresponding master equation are known.  
The  physical origin of these transition rates is a 
pure quantum mechanical phenomenon of tunneling. However, the reduced description of the
system dynamics by means of master equation corresponds to the Markov classical stochastic process. 
The mutual quantum correlations between subsequent tunneling events are not taking into 
account in such approach.  It is a price one has to pay to go from the quantum-mechanical
to classical description. On the other hand, it enables us to use classical probability
methods to derive the FCS.


  In the following we will partially use the notation of the book~\cite{vanKampen}. 
We assume that for each outcome of the experiment one can put into correspondence 
a random set of points at the time axis, obeying the condition
\begin{equation}
 +T/2 > \tau_1 > \tau_2 > \dots > \tau_{s-1} > \tau_s > -T/2
\label{Times} 
\end{equation}
Here $\tau_i$ is the instantaneous moment of $i$'s transition in the system,
and $s=0,1,\dots$ is a non-negative integer, corresponding to the total number 
of transitions during the time of measurement $T$. In the end of calculation
this time should be understood as infinitely large, $T\to+\infty$, so that in average
$\bar s \gg 1$. Given this set of points we introduce the elementary random event 
$\zeta_s=(\tau_1, k_1, \sigma_1; \dots ; \tau_s, k_s, \sigma_s)$. It corresponds to
the experimental outcome, when at time $\tau_i$ the tunneling happens
through the junction $k_i$, $\sigma_i = \pm 1$ being the direction  of the transition.
The events $\zeta_s$ constitute the set $\Omega$ of all possible experimental outcomes. 

At the next step one should define the measure (or the probability) $d\mu(\zeta)$
at the set $\Omega$. For this purpose we may very generally introduce the
sequence of non-negative probabilities
$Q_s(\{\tau_i, k_i, \sigma_i\})\equiv  
Q(\tau_1,k_1,\sigma_1; \dots ;\tau_s, k_s, \sigma_s)\ge 0$
defined at the domain~(\ref{Times}) so that
\begin{equation}
 d\mu(\zeta) = Q_0 + 
 \sum_{s=1}^{+\infty} \sum_{\{k_i,\,\sigma_i\}}
Q_s(\{\tau_i, k_i, \sigma_i\})  d\tau_1 \dots d\tau_s 
\label{measure} 
\end{equation}
The functions $Q$ are normalized according to the condition
\begin{gather}
 \int_{\Omega}d\mu(\zeta) \equiv Q_0 + \nonumber \\  \sum_{s=1}^{+\infty}
\sum_{\{k_i,\,\sigma_i\}}\quad
\idotsint\limits_{T/2>\tau_1>\dots>\tau_s>-T/2}
Q_s(\{\tau_i, k_i, \sigma_i\}) \prod_{i=1}^s d\tau_i  = 1
\label{Norm}
\end{gather}
Each term in Exp.~(\ref{measure}) corresponds to the probability 
of an elementary event $\zeta_s$.

  To complete the preliminaries it is also necessary to define the concept of 
a stochastic process. Mathematically speaking, it can be any integrable function
$\check A(t) \equiv A(t,\zeta)$ defined at the space of all experimental outcomes
$\Omega$ and parametrically depending on time. It is sometimes convenient to omit the explicit 
$\zeta$ dependence. We will use a "check" in this case to stress that the quantity in 
question is a random variable.
Each stochastic process $A(t,\zeta)$ generates the sequence 
of time dependent functions
$\Bigl\{ A_0(t), \, A_1(t,\tau_1,k_1,\sigma_1), \dots ,\, 
A_s(t,\{\tau_i, k_i, \sigma_i\}) \Bigr\}$
Its average ${\langle \check A(t) \rangle}_\Omega$ over the space $\Omega$ is defined as
\begin{gather}
{\langle \check A(t) \rangle}_\Omega = \int_{\Omega} A(t,\zeta) d\mu(\zeta) \equiv 
A_0(t) Q_0 + 
\sum_{s=1}^{+\infty}\sum_{\{k_i,\, \sigma_i\}} \label{Average} \\
\idotsint\limits_{T/2>\tau_1>\dots>\tau_s>-T/2}
A_s(t,\{\tau_i, k_i, \sigma_i\}) 
Q_s(\{\tau_i, k_i, \sigma_i\}) \prod_{i=1}^s d\tau_i  
\nonumber
\end{gather}
The analogous prescription should be used, for instance, to define the correlations 
${\langle \check A(t_1) \check B(t_2) \rangle}_\Omega$
between any two stochastic processes.  

   For the subsequent analysis we define the random process ${\check I^{(k)}(t)}$, 
corresponding to the classical current through the external junction $k\le N$.
It is given by the sequence of the shot $\delta$-pulses in time
\begin{equation}
 I^{(k)}(t, \zeta_s) = \sum_{i=1}^{s} e\sigma_i\, \delta(t-\tau_i) \delta(k-k_i)
\label{Current}
\end{equation}
where $\sigma_m$ is included to take into account the direction of the jump
and $\delta(k-k_i)\equiv \delta_{k,k_i}$ is the Kronecker $\delta$ symbol. Given this
defintion at hand, we introduce the generating functional  $S[\{\chi_i(t)\}]$ 
depending on $N$ counting fields $\chi_i(\tau)$, each of them associating with 
a given terminal $i$:
\begin{equation}
\exp(-S[\{\chi_i(t)\}]) = {\left\langle \exp\Bigl\{ i \sum_{n=1}^N 
\int\limits_{-\infty}^{+\infty} d\tau \chi_n(\tau)\check I^{(n)}(\tau)/e
\Bigr\} \right\rangle}_\Omega
\label{Action}
\end{equation}
The above average is assumed in the sense of definitions~(\ref{measure}) and (\ref{Average}).
We will refer to $S[\{\chi_i(t)\}]$ as the action. Its evaluation is the main goal
of this section.
The $m$-order functional derivatives of $S[\{\chi_i(t)\}]$ with respect
to $\chi_i$ give the irreducible $m$-order current correlations. 
First derivatives corresponds to the average currents  through terminals; the second
derivatives give the shot noise and noise correlations.
In the low-frequency
limit of current correlations one may use the time-independent counting fields $\chi_i$.
In this case the action $S[\{\chi_i\}]$ allows to express the probability of $N_i$
electrons to be transferred through the terminal $i$ during the time interval $T$
\begin{equation}
P(\{N_i\})= \int_{-\pi}^{\pi} \prod_{i=1}^N \frac{d \chi_i}{2 \pi} e^{-S(\{\chi_i\}) -
i \sum_i N_i \chi_i}.
\label{Probability}
\end{equation}

  The above definitions were rather general, but non-constructive, since the
probabilities $Q$ have not been specified so far.  One has to relate
them with transition rates of the master equation.  We will show below how
this can be achieved with use of Markov property of the system.
Our approach will have much in common with the path integral 
method in  statistical physics. The advantage of such scheme is that
it will readily give an explicit algorithm for calculation the generating 
function~(\ref{Action}).

  To proceed with the construction of measure~(\ref{measure}) we assume that
at initial time $t=-T/2$ the system was in the state $\{n^{(s)}\}$. Starting from this state
it is readily to reconstruct the subsequent time evolution of the charge configuration 
$\{n^{(0)}\}\leftarrow\{n^{(1)}\}\dots \{n^{(s-1)}\}\leftarrow\{n^{(s)}\}$ 
for any given outcome $\zeta_s$. (See Fig.3) 
The choice of $\zeta_s$ specifies that the transition between neighbouring charge
states $\{n^{(i-1)}\}$ and $\{n^{(i)}\}$ happens through the junction $k_i=(\alpha_i,\beta_i)$.
Therefore the sequence $\{n^{(i)}\}$ is given by the relation 
$n^{(i-1)}_{\alpha_i}=n^{(i)}_{\alpha_i}-\sigma_{k_i}$, 
$n^{(i-1)}_{\beta_i}=n^{(i)}_{\beta_i}+\sigma_{k_i}$,  and
$n^{(i-1)}_{\gamma}=n^{(i)}_{\gamma}$ for all $\gamma\ne\alpha_i$ and $\beta_i$.
To write down the probability $Q_s(\{\tau_i, k_i, \sigma_i\})$
we assume that our event $\zeta_s$ constitutes the Markov chain and
rely on two facts: (i) the conditional probability of the system  to survive at state
${n^{(i)}}$  between the times $\tau_{i+1}$ and  $\tau_i$ is proportional to 
$\exp[-\gamma(n^{(i)})(\tau_{i}-\tau_{i+1})]$; (ii) the probability that the transition
occurs through the junction $k_i$ during the time interval $d\tau_i$ at the moment $\tau_i$
is given by $\Gamma_{k_i}^{(\sigma_i)}(n^{(i)})d\tau_i$. 
These arguments suggest that $Q$'s have the form
\begin{gather}
 Q_0  = Z_0^{-1}\exp[-\gamma(n^{(s)})T]\label{Qprob} \\
 Q_s(\{\tau_i, k_i, \sigma_i\})   =  Z_0^{-1}
 \exp[-\gamma(n^{(0)})(T/2-\tau_1)]\Gamma_{k_1}^{(\sigma_1)}(n^{(1)}) \nonumber \\
 \exp[-\gamma(n^{(1)})(\tau_1-\tau_2)]\Gamma_{k_2}^{(\sigma_2)}(n^{(2)})\dots 
 \exp[-\gamma(n^{(s-1)})\nonumber \\ (\tau_{s-1}-\tau_{s})] 
   \Gamma_{k_s}^{(\sigma_s)}(n^{(s)})
 \exp[-\gamma(n^{(s)})(\tau_{s}+T/2)] \nonumber
\end{gather}
where the constant $Z_0$ should be found from the normalization condition~(\ref{Norm}). 
As we will see below, $Z_0=1$. 
In Appendix A we show how the usual description of system dynamics in terms of
master equation can be established on the basis of probabilities~(\ref{Qprob}).

 The above correspondence between the random Markov chain $\zeta_s$ and the probabilities 
$Q$'s~(\ref{Qprob}) gives the key to evaluate the generating function~(\ref{Action}). 
By definition~(\ref{Current}) for any given $\zeta_s$ we have
\begin{gather}
\exp\Biggl\{i\sum_{n=1}^N 
\int\limits_{-\infty}^{+\infty} d\tau \chi_n(\tau) I^{(n)}(\tau, \zeta_s)/e\Biggr\} = 
\prod_{i=1}^s \exp\{i\sigma_i\,\chi_{k_i}(\tau_i)\} \nonumber
\end{gather}
It is assumed here that $\chi_{k_i}=0$ if the transition happens through internal junction 
$k_i>N$ whereas no physically measurable current is generated in this case.
On averaging the latter expression
over all possible configurations $\Omega$ with the weight $d\mu(\zeta)$
we see that the generating function takes the structure
of normalization condition~(\ref{Norm})
\begin{gather}
 Z[\{\chi_i(\tau)\}] \equiv\exp(-S[\{\chi_i(\tau)\}]) = Q_0 + \label{Zchi}\\
\sum_{s=1}^{+\infty}
\sum_{\{k_i,\,\sigma_i\}}\quad
\idotsint\limits_{T/2>\tau_1>\dots>\tau_s>-T/2}
Q_s^{\chi}(\{\tau_i, k_i, \sigma_i\}) \prod_{i=1}^s d\tau_i \nonumber
\end{gather}
Here the $\chi$-dependent functions $Q_s^{\chi}(\{\tau_i, k_i, \sigma_i\})$ are
defined similar to probabilities~(\ref{Qprob}) with the only difference that the rates 
$\Gamma_{k}^{(\sigma)}(n)$ should be substituted by 
$\Gamma_{k}^{(\sigma)}(n)\exp\{i\sigma_k\,\chi_k(\tau_k)\}$ if $k\le N$.

The expression~(\ref{Zchi}) can be written in the more compact and elegant way.
For that we introduce the $\chi$-dependent linear operator $\hat L_{\chi}$ defined as
\begin{eqnarray}
&& \hat L_{\chi}(\tau) = \hat \gamma - \hat \Gamma_{\chi}(\tau), 
\label{Lchi} \\
\hat\Gamma_{\chi}(\tau) &=& \sum_{k=1}^N 
(\hat\Gamma_k^{(+)}e^{i\chi_k(\tau)} + \hat\Gamma_k^{(-)}e^{-i\chi_k(\tau)}) \nonumber \\
&+& \sum_{k=N+1}^L (\hat\Gamma_k^{(+)} + \hat\Gamma_k^{(-)})
\nonumber
\end{eqnarray}
Following the above consideration we have attributed the $\chi$-dependent
factor to each operator $\hat\Gamma_k^{(\pm)}$ ($k=1\dots N$),
corresponding to the transition through the external junction. 
The diagonal part and internal transition operators $\hat\Gamma_k^{(\pm)}$ with $k>N$
remained unchanged. Then we consider the evolution operator
$\hat U_{\chi}(t_1, t_2)$ associated with (\ref{Lchi}). Since  
$\hat L_{\chi}(\tau)$ is in general time-dependent, $\hat U_{\chi}(t_1, t_2)$  is given 
by the time-ordered exponent
\begin{equation}
 \hat U_{\chi}(t_1, t_2) = T_\tau \exp\Bigl\{-\int_{t_2}^{t_1}
 \bigl(\hat\gamma(\tau)- \hat\Gamma_{\chi}(\tau)\bigr)d\tau \Bigr\}
\label{TimeExp}
\end{equation}
The similar construction is widely used in quantum statistics. 
The difference in the present case 
is that at $\chi=0$ the operator $\hat U_{\chi}(t_1, t_2)$ gives
the evolution of probability rather than the amplitude of probability.
 
  With the use of evolution operator (\ref{TimeExp}) the generating function 
(\ref{Zchi}) can be cast into the form
\begin{equation}
Z[\{\chi_i(\tau)\}]  = \langle q_0|\hat U_{\chi}(T/2, -T/2)|n_s\rangle 
\label{UAction}
\end{equation}
To prove it we will argue in the following way. We explore the fact, 
that $\hat\gamma(\tau)$ and $\hat\Gamma(\tau)$ commute under the time-ordered operator 
in Eq. (\ref{TimeExp}) and regard $\hat\Gamma(\tau)$ as a perturbation. 
This gives the matrix element $\langle q_0|\hat U_{\chi}(T/2, -T/2)|n_s\rangle$ 
in the form of series
\begin{gather}
\langle q_0|\hat U_{\chi}(T/2, -T/2)|n_s\rangle  = \langle q_0| e^{-\hat\gamma T}|p_0\rangle +
\label{USeries} \\
\sum_{s=1}^{+\infty}
\langle q_0|T_\tau \exp\Bigl\{-\int_{-T/2}^{T/2} \hat\gamma(\tau) d\tau \Bigr\} 
\sum_{{\bf k}_s {\bm\sigma}_s} \quad \idotsint\limits_{T/2>\tau_1>\dots>\tau_s>-T/2} \nonumber 
\\
\hat\Gamma_{k_1}^{(\sigma_1)}(\tau_1)e^{i\sigma_1\chi_{k_1}(\tau_{k_1})}
 \dots \hat\Gamma_{k_s}^{(\sigma_s)}(\tau_s)e^{i\sigma_s\chi_{k_s}(\tau_{k_s})}|p_0\rangle 
\prod_{i=1}^s d\tau_i
\nonumber
\end{gather}
It follows from the definition (\ref{Qprob}) that each term in this series
corresponds to the function $Q_s^{\chi}(\{\tau_i, k_i, \sigma_i\})$, namely
\begin{gather}
Q_0  =  \langle q_0|e^{-\gamma T}|n_s\rangle \nonumber \\
Q_s^{\chi}(\{\tau_i, k_i, \sigma_i\}) = 
\langle q_0|T_\tau \exp\Bigl\{-\int_{-T/2}^{T/2} \hat\gamma(\tau) d\tau \Bigr\} \\
\hat\Gamma_{k_1}^{(\sigma_1)}(\tau_1)e^{i\sigma_1\chi_{k_1}(\tau_{k_1})}
 \dots \hat\Gamma_{k_s}^{(\sigma_s)}(\tau_s)e^{i\sigma_s\chi_{k_s}(\tau_{k_s})}
 \nonumber
|n_s\rangle 
\end{gather}
Therefore Exp.~(\ref{USeries}) and  (\ref{UAction}) are reduced to the previous 
result (\ref{Zchi}). This finishes the proof.  Note, that
owing to the property (\ref{Equilib}) one has the identity
$Z_0 = \langle q_0|\exp(-T\hat L)|n_s\rangle=1$ at $\chi=0$.  Therefore the probabilities
(\ref{Qprob}) are correctly normalized.

  The way~(\ref{UAction}) the generating function $Z[\{\chi_i(t)\}]$ is written
depends on the initial state  $|n_s\rangle$ of the system. 
It looks artificial and we may show that the choice
of $|n_s\rangle$ does not affect the final results. It is naturally to assume 
that $\chi_k(t)\to 0$ when $t\to -\infty$. i.e. physically speaking, the 
measurement is limited in time. To be specific one may suppose that 
$\chi_k(t)=0$ when $-T/2 < t < -T/2+\Delta t$. If the time interval $\Delta t$
is sufficiently large as compared with the typical transition time $\Gamma^{-1}$
then the system will reach the steady state during this period of time. It follows
directly from the fact that $\exp(-\hat L\,\Delta t )|n_s\rangle \to |p_0\rangle$ 
when $\Delta t \gg \Gamma^{-1}$. Thus one can substitute 
$|n_s\rangle$ to $|p_0\rangle$ in Exp.~(\ref{UAction}). Assuming also the 
limit $T\to \infty$, we arrive to the main result of this section
\begin{equation}
 \exp(-S[\{\chi_i(t)\}]) = \langle q_0| T_\tau \exp\Bigl\{-\int\limits_{-\infty}^{+\infty}
 \hat L_{\chi}(\tau) d\tau \Bigr\}|p_0\rangle
\label{TimeS}
\end{equation}
We see that the generating function can be written in the form of the averaged evolution 
operator which corresponds to the modified operator $\hat L_\chi$, depending on the counting 
fields $\chi_i(\tau)$. 

In the rest of the paper we will deal only with the low frequency limit of the current 
correlations, $\omega \ll \Gamma$, and concentrate on the particle statistics 
(\ref{Probability}).
We are interested in the probability $P(\{N_i\})$ of $N_i$ electrons to be transferred through
the corresponding terminal during a large time interval $t_0 \gg \Gamma^{-1}$.
In this case one can put $\chi_k(t) = \chi_k$ when $0\le t\le t_0$ and $\chi_k(t) = 0$
otherwise. The action (\ref{TimeS}) then reduces to the 
\begin{equation}
S(\{\chi_i\}) = t_0 \Lambda_{\rm min}(\{\chi_i\}) 
\label{Lambda}
\end{equation}
where $\Lambda_{\rm min}(\{\chi_i\})$ is a minimal eigenvalue of the operator $\hat L_\chi$.
Thus the problem of statistics in question, provided the transition rates in the system are 
known, is merely  a problem of the linear algebra.

 In the rest of this section we consider the question of the current
conservation in the nodes. For this purpose we associate the
counting fields $\chi_k$ with each line $k=\{\alpha,\beta\}$ of the graph
and in the appropriate way modify the $L_\chi$ operator. Then we define
the classical current operator 
$\hat J^{(k)}_\chi\equiv \hat J^{\{\alpha,\,\beta\}}_\chi$  through each line 
by means of the relation
\begin{equation}
 \hat J^{(k)}_\chi = ie \frac{\partial \hat L_\chi}{\partial \chi_k} \equiv  e\left( 
\hat\Gamma_k^{(+)}e^{i\chi_k} - \hat\Gamma_k^{(-)}e^{-i\chi_k} \right)
\label{Jcurrent}
\end{equation}
Its average value can be found via relation 
\begin{equation}
 I_k(\{\chi_i\}) = \frac{ie}{t_0}\, \frac{\partial S}{\partial \chi_k} = 
\langle q^{(0)}_\chi| \hat J^{(k)}_\chi |p^{(0)}_\chi\rangle
\label{Jchi}
\end{equation}
It follows from the Eq.~(\ref{Lambda}) and the fact that 
$\Lambda_{\rm min}(\{\chi_i\})=\langle q^{(0)}_\chi| \hat L_\chi |p^{(0)}_\chi\rangle $ with
$\langle q^{(0)}_\chi|$ and $|p^{(0)}_\chi\rangle$ being the eigenvectors of the 
$\hat L_\chi$ operator.  
  
  The average physical currents for  $k\leq N$ are  given by
$\bar I_k  = I_k(\{\chi_i\})\big|_{\chi=0}$ 
Expanding the vector notation in~(\ref{Jchi})
one gets the usual relation for the current in the master equation method. The
current is expressed via transition rates and the steady probability distribution $p_0(\{n\})$. 

  We also introduce the particle number operator $\hat n^{\{\alpha\}}$ in each node, 
given by a usual formula
\begin{equation}
 \hat n^{\{\alpha\}} = \sum_{\{n\}}|n\rangle\,n_\alpha\, \langle n|
\end{equation}
Then after few algebra one see that the relation 
\begin{equation}
 \sum_\beta \pm \hat J^{\{\alpha,\,\beta\}}_\chi = - 
e\left[\,\hat n^{\{\alpha\}},\, \hat L_\chi\,\right]
\label{divJ}
\end{equation}
always holds at any node $\alpha$.
Here the summation is going over all nodes $\beta$, connected to $\alpha$.
The choice of the sign in front of each term under the sum
depends on the situation whether the given directed line $k=\{\alpha,\beta\}$ is going
out or coming into the chosen node $\alpha$. The Eq.~(\ref{divJ}) gives the
charge conservation law in the operator language.  On averaging the latter expression
over the steady distribution, $\langle q_0| \dots |p_0\rangle$, and 
using~(\ref{Eigen}) and (\ref{Jchi}) we arrive at the 
conservation law for the $\chi$-dependent currents at each node $\alpha$
\begin{equation}
 \sum_\beta \pm  I_{\{\alpha,\,\beta\}}(\{\chi_i\}) = 0
\label{Jchicons}
\end{equation}
This also ensures the conservation of the physical current in the
model $\sum_k \bar I_k = 0$, where the sum is extended only to the external junctions $k$.
It follows from summing up the relations~(\ref{Jchicons}) over all internal nodes
$\alpha$ and setting $\chi=0$ afterwards.

\end{section}

\begin{section}{Resonant level model}

 In this section we consider the current statistics of the resonant level model.
First we focus on the non-interacting case. Then we apply the general result of the 
section III to the strongly interacting case and compare the statistics in these two 
different regimes. In the end of the section we re-derive the results of the previous
works concerning the shot noise in these systems. 

  Following the definition~(\ref{Lchi}) and the expression for the rates~(\ref{Rates_non_int}),
the $\hat L_\chi$-matrix of the single resonant level model in the non-interacting regime
reads as 
\begin{equation}
\label{L_non_int}
\hat L_{\chi} = \left( 
\begin{array}{cc}
 \Gamma_{1\leftarrow 0}  & -  \Gamma_{0\leftarrow 1}(\chi) \\
 -\Gamma_{1\leftarrow 0}(\chi)  & \Gamma_{0\leftarrow 1} 
\end{array} \right)
\end{equation}
where
\begin{gather}
\label{Rates_chi}
\Gamma_{1\leftarrow 0}(\chi) = \Gamma_L f_L e^{-i\chi_1} + \Gamma_R f_R e^{-i\chi_2}  \\
\Gamma_{0\leftarrow 1}(\chi) = \Gamma_L (1-f_L)e^{i\chi_1} + \Gamma_R (1- f_R)e^{i\chi_2}
\nonumber
\end{gather}
  Calculating the minimal eigenvalue of this matrix one obtains the current statistics
in the form
\begin{gather}
 S(\chi)  = t_0\Bigl\{ \Gamma_L + \Gamma_R -\sqrt{{\cal D}(\chi)}  \Bigr\}\label{SME} \\
{\cal D}(\chi) = (\Gamma_L + \Gamma_R)^2 +  4\Gamma_L\Gamma_R \times \nonumber \\
\left[ f_{(-)}(\epsilon_i)( e^{-i\chi} -1) +  f_{(+)}(\epsilon_i)( e^{i\chi} -1) \right] 
\nonumber
\end{gather}
Here $f_{(-)}(\epsilon_i) =f_L(\epsilon_i)[1-f_R(\epsilon_i)]$, 
$f_{(+)}(\epsilon_i) =f_R(\epsilon_i)[1-f_L(\epsilon_i)]$ and $\chi = \chi_1-\chi_2$.
We have also accounted for the double occupancy of the level by multiplying the result by two. 

  Since the Coulomb blockade phenomenon is completely disregarded in this model, one might
have come to the same result in the framework of the pioneering approach by
Levitov and co-workers~\cite{Levitov}. We will show now that it is indeed the case.

 Following Ref.~\cite{Levitov} the general expression for the current statistics through a single
contact is written as 
\begin{gather}
 S(\chi) = -\frac{t_0}{\pi}\sum_n \int d\epsilon\, {\rm ln}\Bigl\{1+ T_n(\epsilon)\times
\label{SLLL} \\
\Bigl( f_L(\epsilon)[1-f_R(\epsilon)]( e^{-i\chi} -1) + 
f_R(\epsilon)[1-f_L(\epsilon)]( e^{i\chi} -1) \Bigr)  
\Bigr\} \nonumber
\end{gather}
It is valid for any two-terminal geometry provided the region in between two electrodes
can be described by the one-particle scattering approach and the effects of interaction
are of no importance. $T_n(\epsilon)$ is a set of transmission eigenvalues which are in
general energy-dependent. Fermi functions include the effects of applied voltage
and the temperature. In the particular case of a single resonant level the only
resonant transmission eigen-channel $T_r(\epsilon)$ plays the essential role, its energy
dependence is being given by the Breit-Wigner formula
\begin{equation}
 T_r(\epsilon) = \frac{\Gamma_L \Gamma_R}{ (\epsilon - \epsilon_i)^2 + 
(\Gamma_L + \Gamma_R)^2/4}
\end{equation}
Here $\epsilon_i$ denotes a position of a resonant level.  The result~(\ref{SLLL}) is more 
general,
than Exp.~(\ref{SME}). When electrons are supposed to be non-interacting the former
is valid for any temperature. We will show below that one can reproduce
the statistics~(\ref{SME})
on substituting $T_r(\epsilon)$ into the Exp.~(\ref{SLLL}) and assuming the regime
$k_B T\gg \hbar\Gamma$. As it was discussed previously, this is the condition, when the master
equation approach, and hence its consequence~(\ref{SME}),  are valid. 
   
 It is easier to perform the calculation if one first evaluates the $\chi$-dependent
current $I(\chi) = (ie/ t_0)\partial S/\partial\chi$. It reads
\begin{gather}
\label{Ichi}
 I(\chi) = \frac{1}{\pi} \int d\epsilon\, 
\left[ f_{(+)}(\epsilon) e^{i\chi} - f_{(-)}(\epsilon) e^{-i\chi}  \right]
 \times  \\ 
\Bigl\{T^{-1}_r(\epsilon) +  
\left[ f_{(-)}(\epsilon)( e^{-i\chi} -1) + f_{(+)}(\epsilon)( e^{i\chi} -1) \right]  
\nonumber
\Bigr\} 
\end{gather}
In what follows we assume that the resonant level is placed between the chemical
potentials $\mu_{L\{R\}}$ in the leads. 
Since $k_B T\gg \Gamma_{L(R)}$, the main contribution comes from the Lorentz peak
and one can put $\epsilon=\epsilon_i$ in the Fermi functions. 
Therefore we left only with the two poles 
$\epsilon_{1(2)} =\epsilon_i \pm i \sqrt{\cal{D}(\chi)}/2$ 
under the integrand~(\ref{Ichi}). Closing the integration contour in the upper or
lower half-plane we arrive at 
\begin{equation*}
I(\chi) = 2e\,\Gamma_L\Gamma_R
\left[ f_{(+)}(\epsilon_i)e^{i\chi} - f_{(-)}(\epsilon_i) e^{-i\chi}\right]/\sqrt{\cal D(\chi)}
\end{equation*}
On integrating it over $\chi$ one finds for the 
$S(\chi) = (t_0/ie)\int_0^{\chi}I(\chi')d\chi'$ the result~(\ref{SME}) obtained
by means of master equation. Thus, we have verified the correspondence between 
two approaches to statistics in the non-interacting regime.
     
   To proceed we address the strongly interacting regime. In this case
the ($2\times 2$) $\hat L_{\chi}$-matrix is formed with the use of rates~(\ref{RatesU}).
It has a structure similar to Eq.~(\ref{Rates_chi}), provided the $\chi$-dependent rates
are written as
\begin{gather}
\label{RatesUchi}
\Gamma_{1\leftarrow 0}(\chi) = 2\,\Gamma_L f_L e^{-i\chi_1} + 2\,\Gamma_R f_R e^{-i\chi_2}  \\
\Gamma_{0\leftarrow 1}(\chi) = \Gamma_L (1-f_L)e^{i\chi_1} + \Gamma_R (1- f_R)e^{i\chi_2}
\nonumber
\end{gather}
On evaluating the corresponding eigenvalue $\hat L_\chi$ one can write down
the expression for statistics in the strongly interacting limit
\begin{gather}
 S(\chi)  = (t_0/2)\Bigl\{ \Gamma_L[1+f_L(\epsilon_i)] + \Gamma_R[1+f_R(\epsilon_i)] - 
\sqrt{{\cal D}(\chi)} \Bigr\} \nonumber \\
{\cal D}(\chi) = \Bigl\{\Gamma_L[1+f_L(\epsilon_i)] + \Gamma_R[1+f_R(\epsilon_i)]\Bigr\}^2
 +  8\Gamma_L\Gamma_R \times \nonumber \\
\left[ f_{(-)}(\epsilon_i)( e^{i\chi} -1) +  f_{(+)}(\epsilon_i)( e^{-i\chi} -1) \right] 
\label{SMEU}
\end{gather}

  To proceed we consider the shot noise regime $eV\gg k_B T$ and assume that the voltage
is applied to the right electrode as shown in Fig.2.  Then the temperature fluctuations
become non-essential and both statistics~(\ref{SME}) and (\ref{SMEU}) take a rather simple
form
\begin{eqnarray}
 S(\chi)|_{U=0} &=& t_0\Bigl\{ \Gamma_L+\Gamma_R - \nonumber \\
&&\sqrt{(\Gamma_R-\Gamma_L)^2 + 4\Gamma_L\Gamma_R\,e^{-i\chi} }  \Bigr\} 
\label{S_lowT} \\
 S(\chi)|_{U\to\infty} &=& \frac12 t_0\Bigl\{2\Gamma_L+\Gamma_R - \nonumber \\
&& \sqrt{(2\Gamma_L-\Gamma_R)^2 + 8\Gamma_L\Gamma_R\,e^{-i\chi} } \Bigr\} 
\label{S_lowTB}
\end{eqnarray}

\begin{figure}[t]
\begin{center}
\includegraphics[scale=0.35]{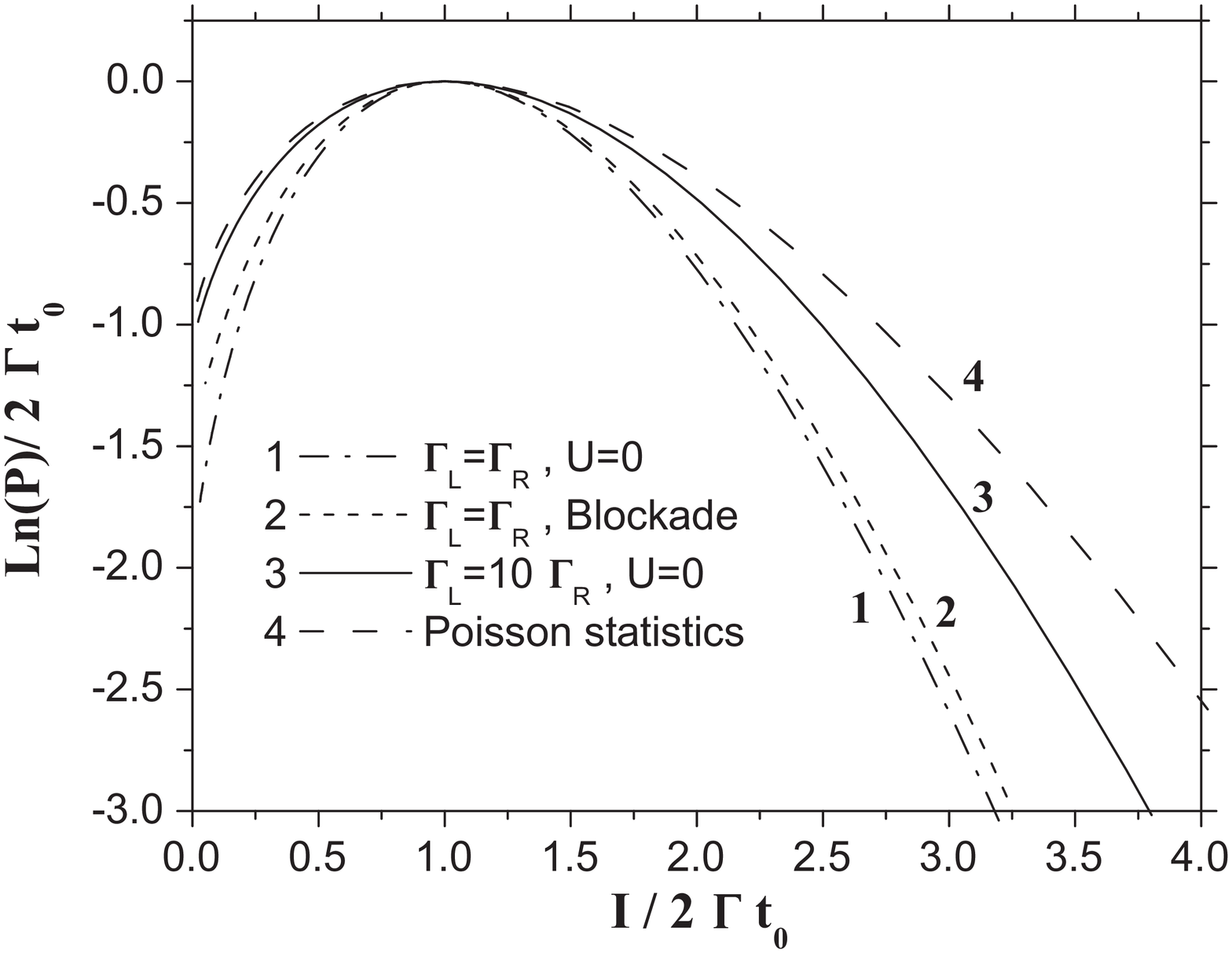}
\caption{ The current statistics through the single resonant level. 
1 and 4 - non-interacting model; 2 - interacting model; 4 - Poisson statistics
}
\end{center}
\end{figure}

  Given the latter expressions at hand one can easily re-derive the known result for the 
average
current $I = (ie/t_0)\partial S/\partial\chi|_{\chi=0}$ and the shot noise power
$S_{\rm shot} = (e^2/t_0)\partial^2 S/\partial\chi^2|_{\chi=0}$ in these models. It is 
conventional to represent $S_{\rm shot}$ in the form $S_{\rm shot} = eIF$, where $F$ is the 
so-called Fano factor. Then one obtains $F=(\Gamma_L^2+\Gamma_R^2)/(\Gamma_L+\Gamma_R)^2$
in the non-interacting regime and $F=(\Gamma_L^2+4\Gamma_R^2)/(\Gamma_L+2\Gamma_R)^2$
in the Coulomb blockade limit~\cite{Struben}.

 As one can see from the Eq.~(\ref{S_lowT}), at low temperatures, the difference of statistics 
in the large $U$ limit from the one in the non-interacting case is an effective suppression of
$\Gamma_R$ rate by a factor of two. 
To find the probability distribution $P(N,t_0)$ one can estimate the 
integral~(\ref{Probability}) by means of steepest descent method. It is applicable in
the given case of low frequency regime $\omega\ll\Gamma$, which we consider, since both
the action $S(\chi)\gg 1$ and the average number of transmitted electrons 
$\bar{N} = \bar{I}t_0/e\gg 1$. Then one has to find the saddle point $\chi_*$ of the
function $\Omega(\chi) = S(\chi)+ i\chi I t_0/e$, which is defined by the equation  
$I = (ie/t_0)[\partial S/\partial\chi]$. It is turned out that $\chi_*$ always lies on 
the imaginary axis. This equation can be regarded as a parametric relation between
$I$ and $\chi_*$, and with the exponential accuracy we obtain the estimation for
the probability $P(I)\sim \exp[-\Omega(\chi_*)]$.

  The results for the statistics~(\ref{S_lowT}) are shown in Fig.~4. The statistics are 
compared
with the Poisson type statistics $S(\chi) = 2\Gamma t_0 [\exp(i\chi)-1]$
with the effective rate given by 
$\Gamma^{-1}=\Gamma_L^{-1}+\Gamma_R^{-1}$. Both statistics (\ref{S_lowT}) approach
the Poisson one, provided the system is strongly asymmetric, $\Gamma_L\gg\Gamma_R$.

\end{section}

\begin{section}{The FCS in the Coulomb blockade quantum dots.}
  
  In this section we discuss the application of the method to the many-terminal
Coulomb blockade quantum dot. The consideration will be limited to the two- and three-terminal
layouts. Our treatment will be mainly numerical, though some analytical
results in the two-terminal setup are also plausible. In the beginning of the section a few
technical  details, which are common for both cases, are given. In particular, we establish
the relation of the FCS approach with the preceding papers, concerning the shot noise
in the conventional two-terminal Coulomb blockade dots. In the following we consider 
the FCS, first for two-terminal, and then for three-terminal configurations.
We will also compare the FCS in the strongly Coulomb blockade limit with our recent
results, concerning the FCS in many-terminal chaotic quantum dot with 
contacts being the tunnel junctions.

\begin{subsection}{General remarks}

  In case of Coulomb blockade island the macroscopic state of the system is characterized
by the excess charge $Q=ne$, which is quantized in terms of electron charge $(-e)$. 
As we have pointed out in the section II, within the "orthodox theory", the charge $Q$ can be 
changed only by $\pm e$ in course of one tunneling event. 
Therefore the master equation connects the given macroscopic
state $n$ only with the neighboring  states $n \pm 1$.  The corresponding rates 
$\Gamma_{n\pm 1\leftarrow n}$ of these transitions are equal to the sum of $N=2$ or $3$
independent probabilities $\Gamma_{n\pm 1\leftarrow n}^{(k)}$ through the different
junctions, those are given by~(\ref{QRates}) and~(\ref{deltaE}). Along with the lines
of section III, in order to find the FCS of the charge transfer through the island, 
we have to modify the rates $\Gamma_{n\pm 1\leftarrow n}$ into $\chi$-dependent
quantities $\Gamma_{n\pm 1\leftarrow n}^\chi$ in accordance with the rule~(\ref{Lchi})
and to evaluate the minimal eigenvalue
$\Lambda_{\rm min}$ of the three-diagonal matrix $\hat L_\chi$ afterwards. In the given
case it is convenient to write down the latter problem as the eigenvalue problem
for the following  set of linear equations
\begin{equation}
 (\Lambda - \gamma_n)p_n + \Gamma_{n \leftarrow n+1 }^\chi p_{n+1} +
 \Gamma_{n \leftarrow n-1 }^\chi p_{n-1} = 0
\label{System}
\end{equation}
where $\gamma_n = \Gamma_{n \leftarrow n-1 } + \Gamma_{n \leftarrow n+1 }$, and 
$\Gamma_{n \leftarrow n\pm 1}^\chi = \sum_{k=1}^N \Gamma_{n\pm 1\leftarrow n}^{(k)} e^{\pm 
i\chi_k}$.
Sign index $\pm$ denotes the outcome (income) of an electron from (to) the island. 

  In general, we have treated the problem~(\ref{System}) numerically. At sufficiently 
low temperatures $k_B T\ll E_c$, which is mainly the case of the following discussion,
the temperature dependence in rates~(\ref{QRates}) is non-essential. Then one can set
$\Gamma_k^{(\pm)}(n) = \Delta E_{n\pm 1\leftarrow n}^{(k)}/(e^2 R_k)$ when
$\Delta E_{n\pm 1\leftarrow n}^{(k)}/(e^2 R_k)\geq 0$ and $\Gamma_k^{(\pm)}(n) = 0$ 
otherwise. Thus defined rates are linear functions in $n$.
The possible set of $\{n\}$, corresponding to nonvanishing rates,
is limited to some interval $n_{\rm min}\le n \le n_{\rm max}$.  Hence Eq.~(\ref{System})
becomes a finite linear problem. At higher temperatures $k_B T \leq E_c$ we have found
that the increase (decrease) of both $n_{\rm min}$ and $n_{\rm max}$ to extra $7\div 8$
states gives the results up to $10^{-15}$ degree of accuracy in course of the numerical 
procedure.

  The matrix $\hat L_\chi$ of Eq.~(\ref{System}) is non-Hermitian. This fact may cause
an instability in the numerical algorithm when the range $[n_{\rm min}, n_{\max}]$ is
large. However, in most practical cases this problem can be
circumvented by transforming $\hat L_\chi$ to Hermitian form. First we note, that one only
needs to work with pure imaginary counting fields $\chi_k$, as long
as the probability $P(\{N_i\},t_0)$ is estimated in the saddle point approximation.
(See the discussion in the end of the section III.) Hence the rates
$\Gamma_{n \leftarrow n\pm 1}^\chi$ become the positive real numbers. Then we can
apply the linear transformation 
$p'_n = A_n p_n$.  It leads to the rates in the new gauge 
${\Gamma'}_{n \leftarrow n\pm 1}^{\chi} = A_{n+1} \Gamma_{n\leftarrow n\pm 1}^\chi A_n^{-1}$. 
The unknown $A_n$'s may be chosen in a way that the symmetry relation
${\Gamma'}_{n \leftarrow n\pm 1}^{\chi} = {\Gamma'}_{n\pm 1 \leftarrow n }^{\chi}$ would hold.
This gives the recurrent relation  $A_{n+1}/A_n = 
(\Gamma_{n \leftarrow n+1}^{\chi}/\Gamma_{n+1 \leftarrow n }^{\chi})^{1/2}$. With the use of 
latter
the Eq.~(\ref{System}) takes the Hermitian form  when it is written in terms of $p'_n$ and 
transformed rates ${\Gamma'}_{n \leftarrow n\pm 1}^{\chi} = 
(\Gamma_{n \leftarrow n \pm 1}^{\chi} \Gamma_{n \pm 1 \leftarrow n }^{\chi})^{1/2}$.
The diagonal term $\gamma_n$ is not affected under this transformation.

 Let us also discuss a useful relation for the shot noise correlations 
$S_{km} = (e^2/t_0)\partial^2 S/\partial\chi_k\partial\chi_m \big|_{\chi=0}$.
One can use the identity 
$\Lambda_{\rm min}(\{\chi_i\})=\langle q^{(0)}_\chi| \hat L_\chi |p^{(0)}_\chi\rangle$ 
in order to express them in terms of eigenvectors $|p_n\rangle$, $\langle q_n|$ 
and eigenvalues $\lambda_n$ of the matrix $\hat L$. With the use of standard algebra
and assuming the normalization $\langle q_n|p_{n'}\rangle = \delta_{n,n'}$ we can cast
$S_{km}$ in the form
\begin{gather}
\label{noise}
 S_{km} = e^2\langle q_0|\hat S^{(k,m)}|p_0\rangle + \\
 e^2\sum_{l>0}\frac{1}{\lambda_l} \left\{ 
 \langle q_0|\hat J_k |p_l\rangle \langle q_l|\hat J_l|p_0\rangle +
 \langle q_0|\hat J_m |p_l\rangle \langle q_l|\hat J_k|p_0\rangle \right\} \nonumber
\end{gather}
where the $\hat J_k$ operator was defined by~(\ref{Jcurrent}) and 
$\hat S^{(k,m)} = \partial^2 \hat L_\chi/\partial\chi_k\partial\chi_m\big|_{\chi=0}$.
Note, that the relation~(\ref{noise}) holds in any basis. One may, in particular, use
it for the basis $p'_n$ discussed above. In this case one must define
the matrix elements of $\hat J_k$ as
$\bigl[\hat J_k\bigr]_{n\pm 1,n} = 
(i\partial/\partial\chi_k) {\Gamma'}_{n \leftarrow n\pm 1}^{\chi}$ and to evaluate 
$\hat S^{(k,m)}$ in the same manner.

  The relation~(\ref{noise}) represents the fact, that the shot noise
correlations are defined by the whole spectrum of the relaxation times 
$\tau^{-1}_k = \lambda_k$ in the system. In the case of two-terminal geometry 
it coincides with preceding results of Ref.~\cite{Korotkov,Hershfield}. 
\end{subsection}

\begin{figure}[t]
\begin{center}
\includegraphics[scale=0.3]{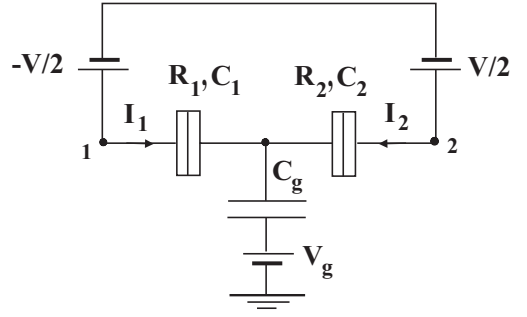}
\caption{The equivalent circuit of the two-terminal Coulomb blockade island.}
\end{center}
\end{figure}

\begin{subsection}{Two-terminal Coulomb blockade island.}
The electrical circuit with a two-terminal Coulomb blockade island is shown in Fig.~5.
The dot is biased such a way, that $V_2=-V_1=V/2$. At low temperatures $k_B T \ll e^2/C_\Sigma$
the $\chi$-dependent rates $\Gamma_{n \leftarrow n\pm 1}^{\chi}$ reads as
\begin{eqnarray}
\label{Rates2T}
\Gamma_{n+1 \leftarrow n}^{\chi} &=& \left[ 
\frac{\tilde C_1 V}{e} - \left(n+ \frac{C_g V_g}{e} + \frac{1}{2}\right) 
\right]\frac{e^{i\chi_2}}{R_2\, C_\Sigma} \\
\Gamma_{n-1 \leftarrow n}^{\chi} &=& \left[ 
\frac{\tilde C_2 V}{e} + \left(n+ \frac{C_g V_g}{e} - \frac{1}{2}\right) 
\right]\frac{e^{-i\chi_1}}{R_1\, C_\Sigma} \nonumber
\end{eqnarray}
where $\tilde C_{1(2)} = C_{1(2)} + C_g/2$ are effective capacitances. 
The gate voltage $V_g$ can be used to control the offset charge $q_0 = C_g V_g$ on the dot.
It can be varied continuously in the range $-e/2\le q_0 \le e/2$.
The resulting dimension of the matrix $\hat L_\chi$ is given by the number of absorbing states
$n_{\rm max} - n_{\rm min}$, where $n_{\rm min}$ ($n_{\rm max}$) is the maximal (minimal)
integer closest to the points $n_{1,2}$ where the rates $\Gamma_{n\mp 1 \leftarrow n}^{\chi}$ 
vanish.

\begin{figure}[b]
\begin{center}
\includegraphics[scale=0.35]{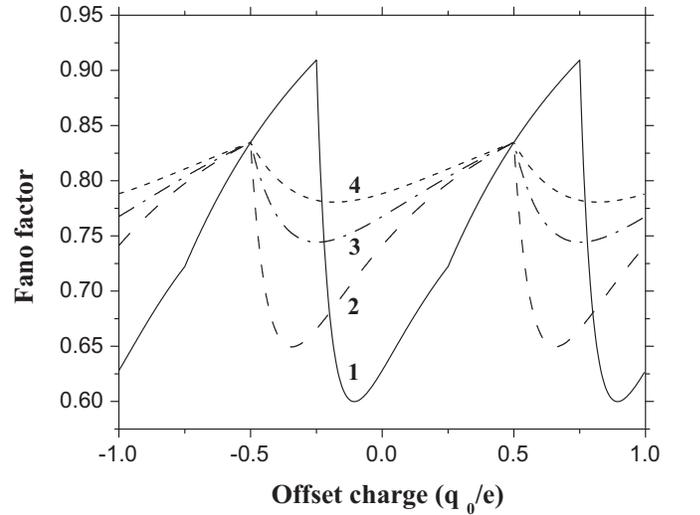}
\caption{ The Fano factor versus the offset charge in the 2~-~terminal Coulomb blockade
dot. The parameters are: $C_1=C_2$, $R_1 = 10\,R_2$, $T =0.01\,e^2/2C_\Sigma$;  
(1) - $V=1.5\,e/C_\Sigma$, (2) - $V=2.0\,e/C_\Sigma$, (3) - $V=4.0\,e/C_\Sigma$, 
(4) - $V=6.0\,e/C_\Sigma$. }
\end{center}
\end{figure}

\begin{figure}[b]
\begin{center}
\includegraphics[scale=0.35]{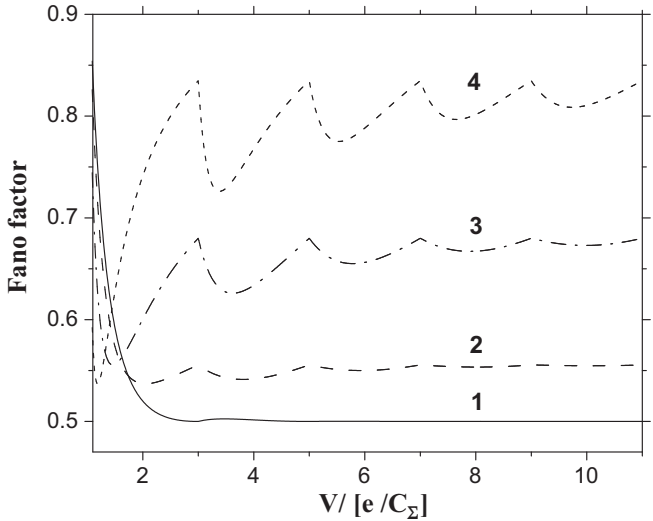}
\caption{ The Fano factor versus the applied voltage in the 2-terminal Coulomb blockade
dot. The parameters are: $C_1=C_2$, $k_B T \ll e^2/2C_\Sigma$,  $q_0 = 0$; 
(1) - $R_1/R_2 = 1$, (2) - $R_1/R_2 = 2$, (3) - $R_1/R_2 = 4$, (4) - $R_1/R_2 = 10$ }
\end{center}
\end{figure}

\begin{figure}[h]
\begin{center}
\includegraphics[scale=0.35]{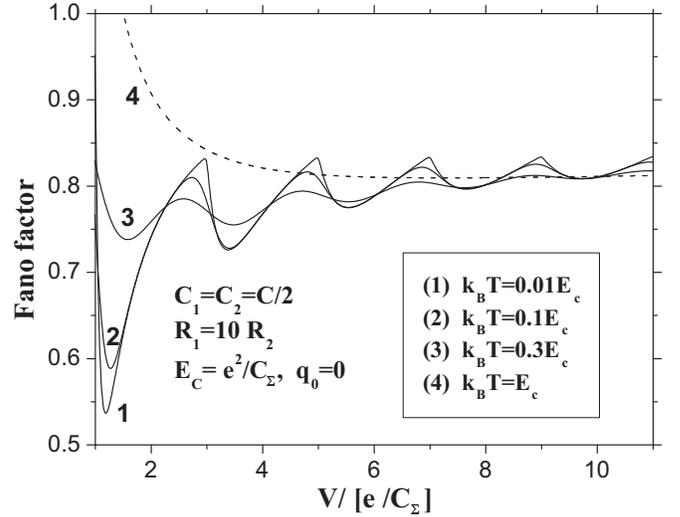}
\caption{ The Fano factor versus the applied voltage at different temperatures. 
Parameters are shown on the plot. Temperature is given in terms of charging energy. }
\end{center}
\end{figure}

  First we briefly consider the voltage dependence of the shot noise in the 
system~\cite{Korotkov, Hershfield}. It was calculated 
with the use of Exp.~(\ref{noise}). The results for the noise-to-current ratio (Fano factor)
are presented in Figs. 6-8. The Coulomb blockade features are strongly pronounced in case
of asymmetric junctions only. In the experimental situation, when the dot is made up 
with the use of $2D$-electron gas in the semiconducting heterostructure, the resistances
of the contacts are much easier to vary than the mutual capacitances. Therefore we have chosen
$\tilde C_1 = \tilde C_2$ and plotted the Fano factor for different values of ratio
$R_2/R_1$ and offset charge $q_0$.  The curves are truncated below the Coulomb blockade
threshold, where the considered "orthodox" theory is not applicable. At high  values of 
the ratio $R_2/R_1$ they exibit the strong characteristic Coulomb blockade oscillations. 
The special points at the voltage dependences occur when either $n_{\rm min}$ or 
$n_{\rm max}$ are changed by 1. At high bias voltages the noise-to-current ratio
saturates to the value $F=(R_1^2+R_2^2)/(R_1+R_2)^2$ independently  of the capacitances $C_k$
and the offset charge $q_0$. (See the discussion below as well.) An increase of a temperature
leads to the smearing of oscillations due to the additional thermal noise.
The above results coincide with those, obtained previously by Hershfiled 
{\it et. al.}~\cite{Hershfield}

   To proceed we turn to the question of the FCS. For the sake of clarity we first present
our recent analytical results for the FCS of the chaotic quantum dot 
with two tunnel junctions, when their resistances  $R_k\ll \pi\hbar/e^2$.
In this case the effects of Coulomb interaction are negligible. Then we trace the differences 
in the FCS, when the dot is placed in the strongly interacting regime $R_k\gg \pi\hbar/e^2$. 

   In the non-interacting limit the action $S(\chi)$ is expressed via the voltage $V$ and the
resistances $R_k$ only~\cite{NazBag}. 
At low temperatures $k_B T \ll eV$ it has a form similar to~(\ref{S_lowT})
\begin{eqnarray}
\label{term2U0}
 S(\chi) &=& \frac{V t_0}{2e}\Bigl\{ R_1^{-1}+R_2^{-1} - \\
&&\sqrt{(R_1^{-1}- R_2^{-1})^2 + 4 (R_1 R_2)^{-1} e^{i(\chi_2-\chi_1)} }  \Bigr\} \nonumber
\end{eqnarray}
It would be completely equivalent to the statistics of the charge transfer by non-interacting
particles through the resonant level if one regards the ratios $\Gamma_{L,R} = V/(eR_{1,2})$
as the effective tunneling rates.  The generating function~(\ref{term2U0}) gives the 
above mentioned value $F=(R_1^2+R_2^2)/(R_1+R_2)^2$ for the Fano factor. 
  
   In the strongly Coulomb blockade limit the action $S(\chi)$, in general,  remarkably
deviates from Exp.~(\ref{term2U0}). Still, there are two exceptions, when $S(\chi)$ resembles
the statistics~(\ref{S_lowT}) and (\ref{term2U0}).  

  The first case occurs at low voltages, slightly
above the Coulomb blockade threshold value, when only one charging state is available
for tunneling. This situation can be easily realized in the asymmetric dot with $R_2\ne R_1$.  
Then mere two states with $n=0$ and $n=1$ are involved and 
$\hat L_\chi$ is reduced to the $2\times 2$ matrix~(\ref{L_non_int}). The only 
difference is that the rates $\Gamma_{L,R}$ contain the voltage dependence as given
by Exp.~(\ref{Rates2T}). Thus the action $S(\chi)$ reproduces the result~(\ref{S_lowT}),
where the rates $\Gamma_{L,R}$ are assumed to be voltage dependent.

   To proceed we describe the second exceptional situation when the action $S(\chi)$
can be found analytically.  Let $n_{1,2}$ be zeros of rates~(\ref{Rates2T}), i.e. 
$\Gamma^\chi_{n\pm 1 \leftarrow n}(n_{1,2}) = 0$.  We now interested in the situation
when both zeroes $n_{1,2}$ simultaneously become integers. This situation may occur
at the limited number of special points $V_k$ at the Coulomb blockade staircase
when the ratio $\tilde C_1/\tilde C_2$ is close to a rational along
with the special choice of the offset charge $q_0$.
(E.g., for the configuration $\tilde C_1=\tilde C_2$,
shown in Figs.~6, 8, this is the case when (i) $q_0=0$, $V_k = (2k+1)e/C_\Sigma$ and (ii)
$q_0 = \pm e/2$, $V_k = 2k (e/C_\Sigma)$ with $k$ being integer.) In this situation
we may show analytically (see Appendix B for the proof), that the action $S(\chi)$ at points 
$V_k$ takes the form similar to the statistics in the non-interacting regime~(\ref{term2U0}).
The only difference that the voltage $V$ has to be substituted to $(|V_k|-e/C_\Sigma)$.
The reduction of $V$ by the amount of the threshold voltage value $e/C_\Sigma$ is thus
the manifestation of the Coulomb interaction.  The given statistics is also valid
as a limit at high voltages $V\gg e/C_\Sigma$.  One may conclude it from the physically
reasonable arguments that the result for the action in this limit should be linear 
function in voltage and must not depend on the capacitance ratio $\tilde C_1/\tilde C_2$.
Hence, the statistics 
is insensitive to the fact, whether $n_{1,2}$ are integers or not. This also
explains the saturation of the Fano factor at Figs. 6-8 to the  non-interacting 
current-to-noise ratio. 

\begin{figure}[b]
\begin{center}
\includegraphics[scale=0.36]{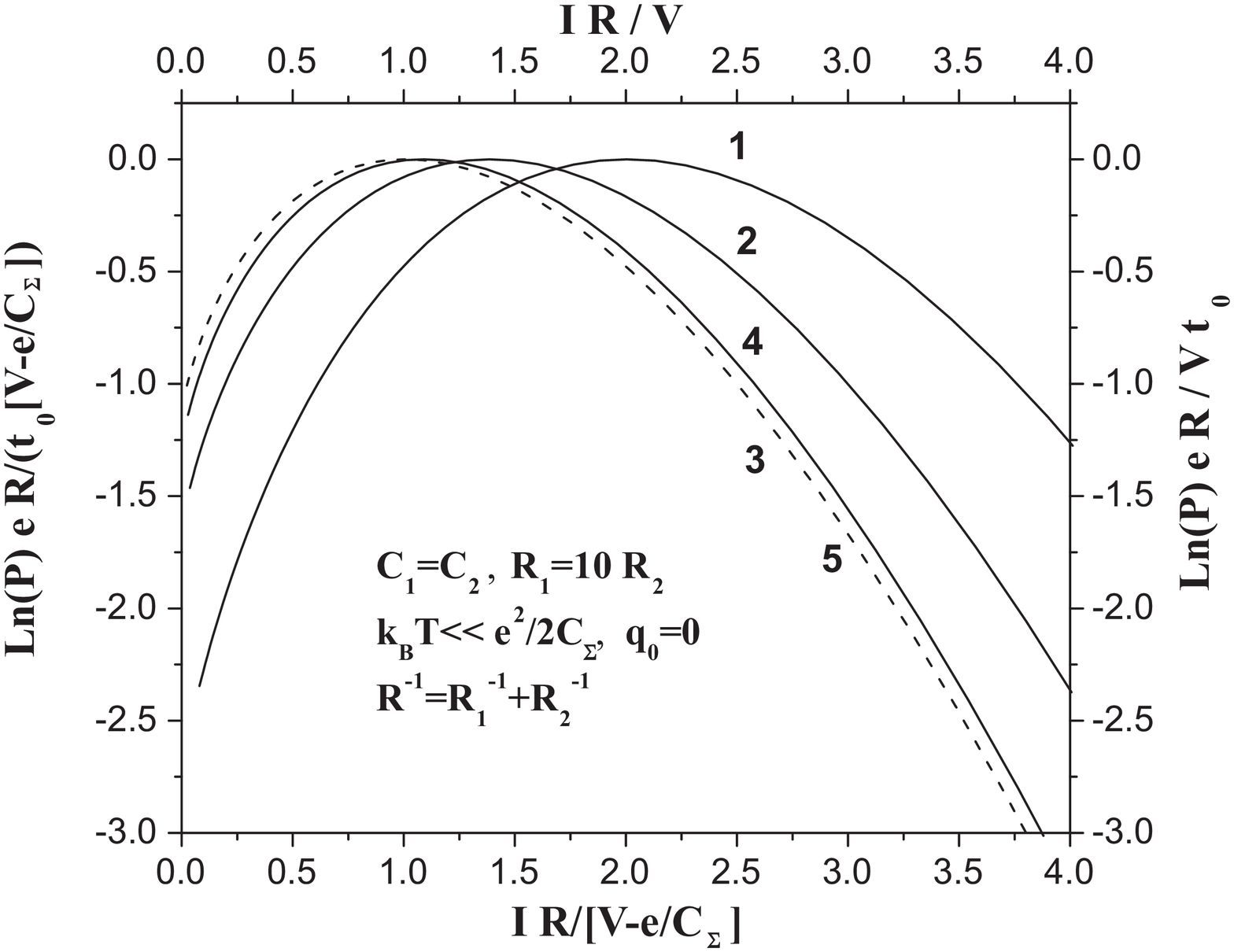}
\caption{ The statistics of current in the two-terminal quantum dot. Parameters are
shown on the plot. Curves 3 and 5 coincide but correspond to different axes. 
1) - $V=1.5\,e/C_\Sigma$, (2) - $V=2.0\,e/C_\Sigma$, (3) - $V=3.0\,e/C_\Sigma$, 
(4) - $V=4.0\,e/C_\Sigma$, (5) - non-interacting regime. }
\end{center}
\end{figure}

   To access the general situation, the whole problem has been treated numerically. We
have evaluated the probability~(\ref{Probability}) in the saddle point approximation
along with the same lines as it was done for the resonant level model. The
Eq.~(\ref{Jchi}) was used as the parametric relation  between the current $I$
and the counting field $\chi=\chi_2-\chi_1$ in the saddle point $\chi_*$. In Fig.~9
we give the example of the logarithm of probability distribution $P(I)$ for the
number of different voltages $V$ and offset charge $q_0=0$. All curves are
normalized by the reduced value of the voltage $(V-e/C_\Sigma)$. For the value
$V=3e/C_\Sigma$ the renormalized logarithm of probability coincides with the one,
obtained from the non-interacting limit~(\ref{term2U0}). We have also plotted the
same statistics in Fig.~4 in the resonant level model for the ratio of 
rates $\Gamma_L/\Gamma_R = 10$, when the interaction effects are disregarded.
We see, that in general
the probability distribution is strongly affected by the Coulomb blockade phenomenon,
as compared to the non-interacting regime. It approaches to the non-interacting
limit only at rather high voltages $V\gg e/C_\Sigma$.

\end{subsection}

\begin{subsection}{Three-terminal Coulomb blockade island}
 
\begin{figure}[t]
\begin{center}
\includegraphics[scale=0.3]{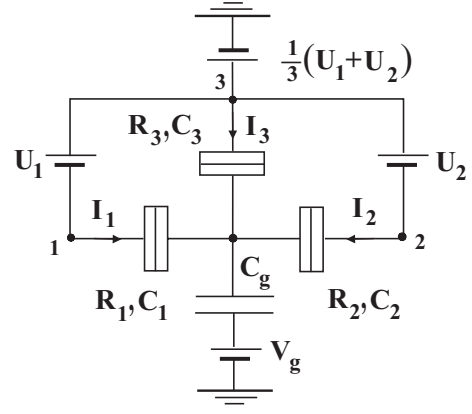}
\caption{The equivalent circuit of the three-terminal Coulomb blockade island.}
\end{center}
\end{figure}

 The electrical circuit with a three-terminal Coulomb blockade island is presented in Fig.~10. 
It is biased by three external voltage sources so that the current, flowing through
the third terminal, would split into the first and the second ones. The voltages $U_{1(2)}$
are used to control the bias between the 3d and the 1st (the 2nd) terminals:
$U_{1(2)}=V_3-V_{1(2)}$. The third terminal is biased at voltage $V_3=(U_1+U_2)/3$ with
respect to the ground. Such type of setup assures the  condition
$V_1+V_2+V_3 = 0$. Hence the gate voltage, $V_g$, can be used as before to control the
offset charge $q_0=C_g V_g$ on the island. As in the previous
subsection we discuss the only low temperature regime $k_B T\ll e^2/C_\Sigma$. 
In what follows it is assumed that $U_2>U_1$.  Then, according  
to Exp.~(\ref{QRates}-\ref{V0n}) and (\ref{Lchi})
the $\chi$-dependent rates of the system are written as follows
\begin{gather}
 \Gamma_{n+1\leftarrow n}^{\chi} = \Gamma_{3}^{(+)}(n)e^{i\chi_3} + 
 \Gamma_{1}^{(+)}(n)\,\theta(m-1/2-n)e^{i\chi_1} \nonumber \\
 \Gamma_{n-1\leftarrow n}^{\chi} = \Gamma_{2}^{(-)}(n)e^{-i\chi_2} + 
 \Gamma_{1}^{(-)}(n)\,\theta(n-m-1/2)e^{-i\chi_1} 
\label{rates3D}
\end{gather}
where
\begin{equation}
\Gamma_{k}^{(\pm)}(n) = a_k^{(\pm)} \mp \left(n+ \frac{C_g V_g}{e} \pm  \frac{1}{2}\right)
\frac{1}{R_k\, C_\Sigma}  
\end{equation}
and
\begin{gather}
a_3^{(+)} = \frac{\tilde C_1 U_1 + \tilde C_2 U_2}{e R_3\, C_\Sigma}, \quad 
a_2^{(-)} = \frac{ (\tilde C_1+ \tilde C_3) U_2-\tilde C_1 U_1}{e R_2\, C_\Sigma} \nonumber \\
a_1^{(\pm)} = \pm \frac{ \tilde C_2 U_2 - (\tilde C_3+ \tilde C_2) U_1}{e R_1\, C_\Sigma} \nonumber
\end{gather}
The point $m$ is determined by the relation $\Gamma_1^{(-)}(m+1/2)=\Gamma_1^{(+)}(m-1/2)=0$ 
($m$ is non-integer in general).
The dimension of the $\hat L_\chi$-matrix is equal to $n_{\rm max}-n_{\rm min}$,
where $n_{\rm max}$($n_{\rm min}$) can be found from the conditions $\Gamma_3^{(-)}(n)\ge 0$
($\Gamma_1^{(+)}(n)\ge 0$). The effective capacitances $C_k$ are defined as $\tilde C_k = C_k + 
C_g/3$

  We can see from the Exp.~(\ref{rates3D}) that there are four elementary processes of charge
transfer in the
system at low temperatures, each of them is being associated with the pre-factor $e^{\pm 
i\chi_k}$. 
The presence of the  exponents $e^{i\chi_3}$ and $e^{-i\chi_2}$  corresponds to the charge 
transfer
from the third terminal into the island and from the island into the second terminal, 
respectively.
Hence, the random current through the 3d (2nd) junctions always have the positive (negative)
sign.  Two factors $e^{\pm\chi_1}$ stems from the charge  transfer
through the first junction in the direction 
either from the island into the first contact or vice versa. Therefore
the random current $I_1$ is able to fluctuate in both directions.

  Let us first consider the shot noise correlations in the system. For that it is useful
to introduce the ($3\times 3$) matrix $F$ with elements $F_{km} = S_{km}/eI_\Sigma$,
where the current correlations $S_{km}$ are given by~(\ref{noise}) and 
$I_\Sigma = \sum_{i=1}^3 |I_i|$. The matrix $F$ is a generalization of the Fano factor for the
multiterminal system. It is symmetric and obeys the relation $\sum_{i=1}^3 F_{ik}=0$. 
It follows from the general law of the current conservation in the system. For the 
considered 3-terminal dot we have found that the cross-correlations $F_{km}$ ($k\ne m$)
are always negative for any set of parameters.

\begin{figure}[b]
\begin{center}
\includegraphics[scale=0.35]{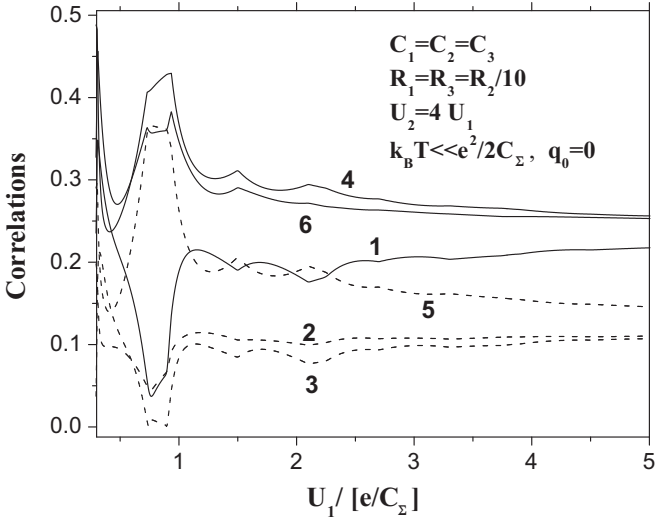}
\caption{ The matrix $F$ of auto- and cross- shot noise correlation versus voltage $U_1$
for the 3-terminal quantum dot setup.
Parameters are shown on the plot.
(1) - $F_{11}$, (2) - $|F_{12}|$, (3) - $|F_{13}|$, 
(4) - $F_{22}$, (5) - $|F_{23}|$, (6) - $F_{33}$.  }
\end{center}
\end{figure}

\begin{figure}[h]
\begin{center}
\includegraphics[scale=0.33]{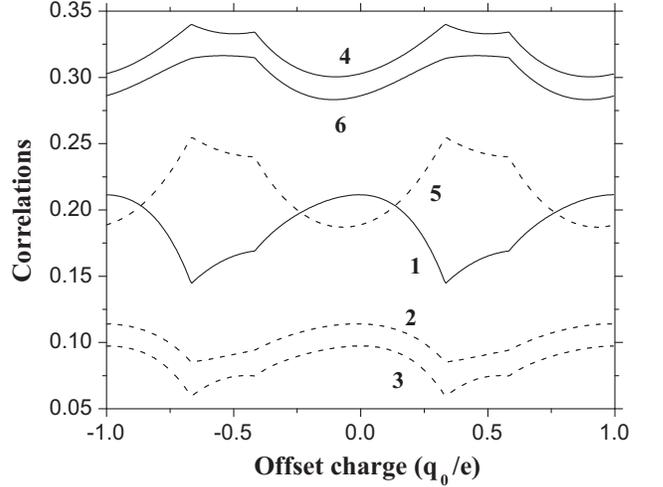}
\caption{ The matrix $F$ of auto- and cross- shot noise correlation versus the offset charge
for the 3-terminal quantum dot setup. Parameters are the same as on the Fig.~11. The voltage
$U_1 = U_2/4 = 1.25 e/C_{\Sigma}$.
(1) - $F_{11}$, (2) - $|F_{12}|$, (3) - $|F_{13}|$, 
(4) - $F_{22}$, (5) - $|F_{23}|$, (6) - $F_{33}$.  }
\end{center}
\end{figure}

 In Fig.~11 we give the illustrative example of the voltage dependence of the shot noise
correlations $F_{km}$ for the certain choice of parameters.  (For the cross-correlations
the modulus $|F_{km}|$ are given.) As in the 2-terminal case
the Coulomb blockade features are strongly  pronounced only for the asymmetric setup.
The results in Fig.~10 corresponds to $R_1=R_3=R_2/10$, $\tilde C_1 = \tilde C_2 =\tilde C_3$
and $U_2/U_1=4$. The latter ratio of voltages has been chosen on the basis of arguments that 
for a given value of resistances $R_k$ it would split the
average current $I_3$ into two equal currents $|I_1|=|I_2|=I_3/2$ provided one could
apply the usual linear Kirchgoff rules to this circuit. In Fig.~12 the dependence of the
shot noise correlations on the offset charge is shown for the same set of parameters and 
the value of $U_1=1.25\,e/C_\Sigma$. The special points of both these 
dependences occur when either $n_{\rm min}$, $n_{\rm max}$ or the integer part
of $m$ are changed by $\pm 1$. As the result we observe multi-periodic Coulomb blockade
oscillations in the offset charge dependences in contrast to the single periodic
oscillations in the two-terminal case.

\begin{figure}[t]
\begin{center}
\includegraphics[scale=0.36]{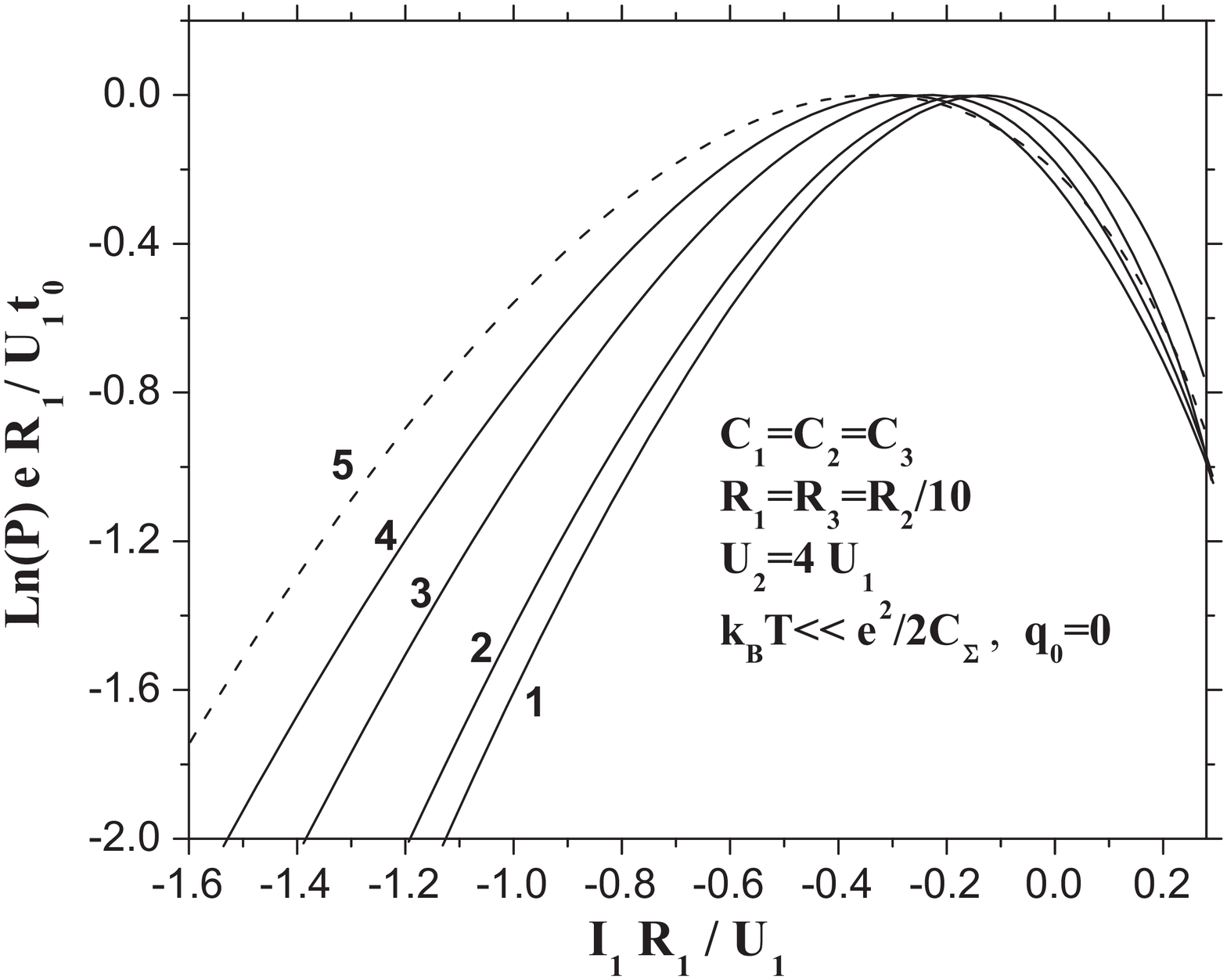}
\caption{ The logarithm of current distribution $\ln P(I_1, I_2)$ in the 3-terminal quantum dot 
as
a function of current  $I_1$, under condition $I_2=\langle I_2 \rangle$. 
Parameters are shown on the plot. (1) - $U_1=1.25\,e/C_\Sigma$, (2) - $U_1=2.0\,e/C_\Sigma$, 
(3) - $U_1=4.0\,e/C_\Sigma$, (4) - $U_1=10.0\,e/C_\Sigma$; curve (5) corresponds to the
non-interacting regime. }
\end{center}
\end{figure}

\begin{figure}[b]
\begin{center}
\includegraphics[scale=0.36]{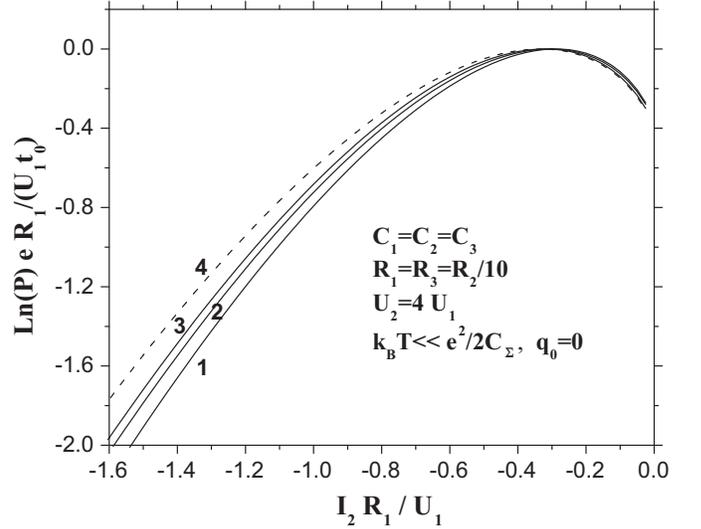}
\caption{ The logarithm of current distribution $\ln P(I_1, I_2)$ in the 3-terminal quantum dot 
as
a function of current  $I_2$, under condition $I_1=\langle I_1 \rangle$. 
Parameters are the same as in Fig.~13. (1) - $U_1=1.25\,e/C_\Sigma$, (2) - 
$U_1=2.0\,e/C_\Sigma$, 
(3) - $U_1=4.0\,e/C_\Sigma$, curve (4) corresponds to the non-interacting regime. }
\end{center}
\end{figure}

\begin{figure}[t]
\includegraphics[scale=0.31]{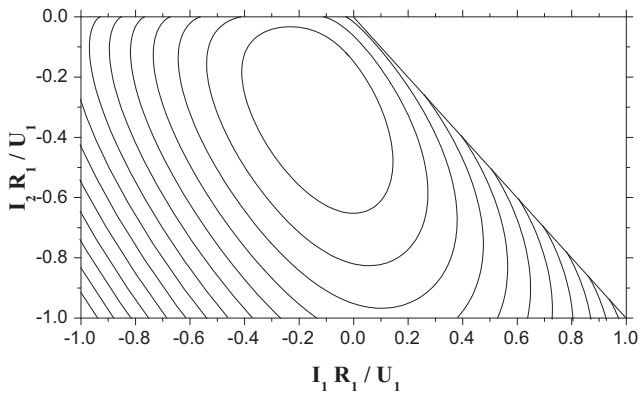}
\caption{The contour maps of the current distribution $\log[P(I_1,I_2)]$ in
the 3-terminal Coulomb blockade dot. Parameters are the same as in Fig.~13.
$U_1=U_2/4.0 = 1.25\,e/C_\Sigma$. }
\end{figure}

\begin{figure}[t]
\includegraphics[scale=0.31]{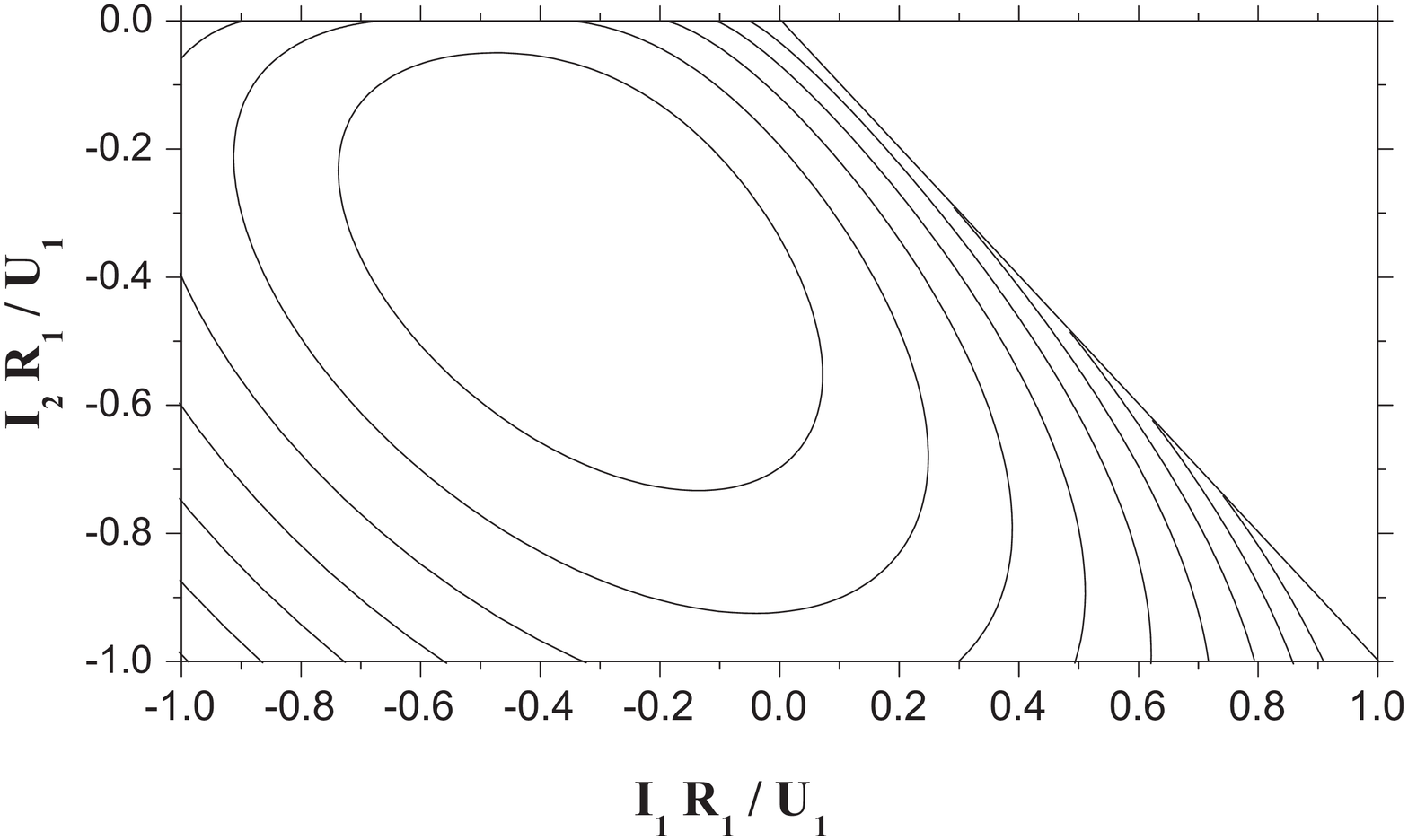}
\caption{The contour maps of the current distribution $\log[P(I_1,I_2)]$ in
the 3-terminal chaotic  quantum dot with tunnel contacts. 
$U_1=U_2/4.0 = 1.25\,e/C_\Sigma$. }
\end{figure}

  We now proceed with the consideration of the FCS. As before, the action $S(\{\chi_i\})$
has been calculated with the use of (\ref{Lambda}) and afterwards the probability
(\ref{Probability}) has been estimated by means of the steepest descent method. 
The difference with the two-terminal geometry is that three currents $I_k$ and three 
counting fields $\chi_k$ are now involved. Due to the current conservation
$\sum_k I_k$ = 0  for any plausible fluctuation, only two  currents 
are independent and thus the action $S(\{\chi_i\})$ depends on the differences
$\chi_{ij} = \chi_i-\chi_j$ only. In what follows we have chosen $I_1$ and $I_2$ as
the independent variables to plot the logarithm of probability $\ln P(I_1,I_2)$. 
With the exponential accuracy it is given by $\ln P(I_1,I_2) \sim e^{-\Omega(\chi^*)}$,
where $\chi^*$ is a saddle point of the function  
$\Omega(\chi) = S(\chi)+ i\chi_1 I_1 t_0/e + i\chi_2 I_2 t_0/e $. The
results for $\ln P(I_1,I_2)$ are shown in Figs.~13, 14, and 15. From the contour map
on Fig.~15 we see that $P(I_1,I_2)$ is non-zero in the quarter $I_1<0$, $I_2<0$
of a current plain $(I_1, I_2)$ and  in
the region $I_1\le |I_2|$ belonging to the quarter $I_1>0$, $I_2<0$. This range of plausible
current fluctuations results from the $\chi$-dependence of rates (\ref{rates3D})
and the restriction $\sum_k I_k$ = 0. As it was discussed above, any current
fluctuation obeys the condition $I_2<0$ and $I_3>0$. There is no additional
restriction to the latter when $I_1<0$. On the other hand, it follows from the current
conservation, that  $I_1=|I_2|-I_3\le |I_2|$ in the region $I_1>0$.

  It is also worth to compare the above results with the FCS in the 3-terminal
chaotic quantum dot when its contact are tunnel junctions with resistances 
$R_k^{-1} \gg {e^2}/\pi\hbar$. In this limit the effects of interaction are 
negligible and electrons are scattered independently at different energies. 
Provided $U_2>U_1$ the generating function $S(\{\chi_i\})$ in the given case is a
sum of the two  independent processes
\begin{equation}
 S(\chi_1,\chi_2,\chi_3) = S_1(\chi_1,\chi_2,\chi_3) + S_2(\chi_1,\chi_2,\chi_3) 
\label{SNoninter}
\end{equation}
Here 
\begin{gather}
 S_1(\chi_1,\chi_2,\chi_3) = \frac{U_1 t_0}{2e}\Bigl\{ G_1 + G_2 + G_3  - \nonumber \\ 
\sqrt{ (G_1 + G_2 - G_3 )^2 + 
4 G_3 e^{i\chi_3} ( G_1 e^{-i\chi_1} + G_2 e^{-i\chi_2} ) } \Bigr\} \nonumber \\
 S_2(\chi_1,\chi_2,\chi_3) = \frac{(U_2 - U_1) t_0}{2e}\Bigl\{ G_1 + G_2 + G_3  - \nonumber \\ 
\sqrt{ (G_1 + G_3 - G_2 )^2 + 
4 G_2 e^{-i\chi_2} ( G_1 e^{i\chi_1} + G_3 e^{i\chi_3} ) } \Bigr\} \nonumber 
\end{gather}
and $G_k = R_k^{-1}$ are the conductances  of the junctions. 

  The logarithm of probability $\ln P_0(I_1, I_2)$, evaluated with the use of 
statistics~(\ref{SNoninter}), is shown by the dashed line in addition to the previous 
curves in Figs.~13, 14. 
Its contour map for the same values of parameters is also separately presented
in Fig.~16.  The maximum of $\ln P_0(I_1, I_2)$, as expected, occur at 
$\bar I_1 = \bar I_2 = U_1/3 R_1$.  

  The two main conclusion which one can derive on
comparing the statistics in the two limiting regimes (strongly interacting in case of the
Coulomb blockade quantum dot and the non-interacting one in the open chaotic
quantum dot) are as follows. First, we see, that in spite of the different regimes,
the qualitative dependence of probabilities versus the currents is similar for both
statistics. Second, we may conclude, that the relative probability of the 
big current fluctuations in the 
Coulomb blockade limit is suppressed with respect to the situation in the 
non-interacting regime.  This is a characteristic feature of both two- and 
three-terminal quantum dots. This behavior is rather natural. It stems from the
fact, that any big current fluctuation in Coulomb blockade dot is related with the
large accumulation (or depletion) of the charge on the island. The latter costs
the extra electrostatic energy and hence the probability of such fluctuation is
drastically decreased. The suppression of the average current due to the
presence of the Coulomb gap in the interacting system is the particular example
of this behavior.

\end{subsection}

\end{section}

\begin{section}{Conclusions}
  To conclude, in the present paper we have developed the constructive scheme to evaluate
the FCS of charge transfer in the Coulomb blockade systems. This scheme is rather general
and universal and is applicable to any strongly interacting system, provided
the latter can be described classically in the framework of the master equation approach. 
The method proposed consists in the transformation of the initial linear operator 
$\hat L$ of the master
equation into the auxiliary $\chi$-dependent linear operator $L_{\chi}$. Each non-diagonal
term of this new operator, associated with the particular transition in the system, is modified
by the exponential prefactor $e^{\pm i\chi_k}$ in order to take into account the electron jump
through the junction $k$ during the tunneling event. The generating function of the 
charge transfer through the whole system is then proportional to the minimal eigenvalue of 
the operator $L_\chi$.

   We have applied this scheme to study the FCS in two different systems.
For a generic case of a single resonant level model we have established the equivalence 
of the new method with the scattering approach to the FCS, when the particles in
the system are non-interacting.

   Afterwards we have considered the FCS and the shot noise correlations in the two- and 
three-terminal Coulomb blockade quantum dots. 
The consideration was limited to the temperature regime when the orthodox theory of
Coulomb blockade phenomenon is applicable. For the case of two-terminal dot we have
re-established all
the known results for the shot noise in this system. In the three-terminal case we have shown 
that the auto- and cross- shot noise correlations exhibit the characteristic 
Coulomb blockade oscillations as the functions of the applied voltages and the offset charge. 
We have considered the question of the FCS as well. 
In general situation we evaluated the probability distribution numerically. However,
at some special values of parameters in two-terminal dot we have managed 
to find FCS analytically. 
In these exceptional cases the FCS resembles the statistics of the charge transfer
through the single resonant level.
Then we compared the statistics in the Coulomb blockade dots with our previous results 
concerning an open dot with two and three terminals. We found that the 
Coulomb interaction suppresses the relative probability of the 
big current fluctuations.
\end{section}

\begin{acknowledgments}
This work is a part of the research program of the "Stichting voor
Fundamenteel Onderzoek der Materie"~(FOM), and we acknowledge the financial
support from the "Nederlandse Organisatie voor Wetenschappelijk Onderzoek"~(NWO). 

\end{acknowledgments}
   
\appendix

\begin{section}{}

In this Appendix we show that the construction of the probability measure
on the basis of Markov chains $\zeta_s$, which was used to derive the main result
of section III, leads to the usual description of the system dynamics in terms of master
equation. This correspondence is achieved in the standard way of probability theory by 
introducing the stochastic process $\check n(t)$ corresponding to the
island charge at a given time $t$
\begin{equation}
 n(t,{\zeta}_s) = n_s + \sum_{i=1}^{s} \sigma_i\,\theta(t-\tau_i)
\end{equation}
Similarly one can consider the random number of electrons $\check n^{(k)}(t)$ transferred 
through the junction $k$ after $t\ge -T/2$
\begin{equation}
 n^{(k)}(t,{\zeta}_s) = \sum_{i=1}^{s} \sigma_i\,\theta(t-\tau_i)\delta(k-k_i)
\end{equation}
The random variables $\check n(t)$ and $\check n^{(k)}(t)$ are subjected to the
relations
\begin{gather}
n(t,{\zeta}_s) = n_s + \sum_{k=1}^N  n^{(k)}(t,{\zeta}_s) \nonumber \\
I^{(k)}(t,{\zeta}_s) = e\frac{\partial}{\partial t} n^{(k)}(t,{\zeta}_s)
\end{gather}

After that we can introduce the probability distribution $P(n,t)$ and
the joint probability distribution $P(n_1, t_1; n_2, t_2)$ ($t_1\ge t_2$) of the
process $\check n(t)$ 
\begin{gather}
 P(n,t)  = \int_{\Omega} \delta(n-n(t,\zeta)) d\mu(\zeta) \label{Ps} \\
 P(n_1,t_1; n_2, t_2)  = \int_{\Omega} \delta(n_1-n(t_1,\zeta)) \delta(n_2-n(t_2,\zeta)) 
d\mu(\zeta) \nonumber
\end{gather}
Their ratio $P(n_1, t_1 |\,n_2, t_2) = P(n_1,t_1; n_2, t_2)/P(n_2,t_2)$ gives the
conditional probability to find the system at state $n_1$ at time $t_1$, 
given that at time $t_2$ it was at state $n_2$. The integrals~(\ref{Ps}) can
be efficiently evaluated along with the same reasoning
as we have used to prove the normalization condition. As the result one ends up with 
\begin{gather}
 P(n_1, t_1 |\,n_2, t_2) = \langle n_1|\hat U(t_1, t_2) | n_2\rangle 
\end{gather}

The latter expression is the usual way to describe the system in terms
of master equation. The conditional probability $P(n_1, t_1 |\,n_2, t_2)$  
regarded as a function of $n_1$ and $t_1$ obeys this equation with the
initial condition  $P(n_1,t_1)=\delta_{n1,n_2}$ at $t_1=t_2$.

\end{section}

\begin{section}{}
This appendix contains the derivation of the action $S(\chi)$ at low temperatures
at some special points $V_k$ in the Coulomb blockade staircase in the two-terminal
dot. We introduce the notation $\Gamma^{(\pm)}_\chi(n) = \Gamma_{ n\pm\leftarrow n}^\chi$,
that enables to write down Eq.~(\ref{System}) in the form
\begin{equation}
 (\Lambda - \gamma_n) p_n + \Gamma_\chi^{(-)}(n+1) p_{n+1} + \Gamma_\chi^{(+)}(n-1) p_{n-1} = 0
\label{System2}
\end{equation}
If all $\chi_k = 0$ then the stationary solution of this equation, corresponding to
$\Lambda = 0$, satisfies the detailed balance condition
$p_{n+1}\Gamma_0^{(-)}(n+1) = p_n\Gamma_0^{(+)}(n)$. In general situation, when $\chi_k\ne 0$,
one may try to resolve~(\ref{System2}) making use of the substitution 
\begin{equation}
 \frac{p_{n+1}}{p_n} = \frac{\Gamma_\chi^{(+)}(n)}{y\, \Gamma_\chi^{(-)}(n+1) }, \quad
 \frac{p_{n-1}}{p_n} = \frac{y\, \Gamma_\chi^{(-)}(n)}{\Gamma_\chi^{(+)}(n-1) } 
\label{subst}
\end{equation}
with unknown constant $y$ to be found. This reduces the difference equation~(\ref{System2})
to the relation 
\begin{equation}
 (\Lambda - \gamma_n)  + \Gamma_\chi^{(-)}(n) y + \Gamma_\chi^{(+)}(n) y^{-1} = 0.
\label{Relation}
\end{equation}
Here $\gamma_n = \Gamma_0^{(-)}(n) + \Gamma_0^{(+)}(n)$,  and $\Gamma_\chi^{(\pm)}(n)$ are
linear functions in $n$, given by Exp.~(\ref{Rates2T}). Then one might find the two unknown
$y$ and $\Lambda_\chi$ on comparing the constant and linear in $n$ terms in 
relation~(\ref{Relation}). It yields
\begin{gather}
 y =  \left( R_1^{-1}- R_2^{-1} +  \sqrt{\cal{D}(\chi)} \right)/2 R_1^{-1} e^{-i\chi_1} 
\nonumber \\
 \Lambda(\chi) =  \frac{1}{2e}(V-e/C_\Sigma)\left( R_1^{-1} + R_2^{-1} -  \sqrt{\cal{D}(\chi)} 
\right) \\
{\cal D}(\chi) = (R_1^{-1}- R_2^{-1})^2 + 4 (R_1 R_2)^{-1} e^{i(\chi_2-\chi_1)} \nonumber
\end{gather}
It looks like we have found in such a way the required solution. However, it is
not valid at all possible values of parameters.  The matter is that in the most general situation
the expressions~(\ref{Rates2T}) are not correct at points $n=n_{\rm min}$ and $n=n_{\rm max}$.  
One have to set $\Gamma_\chi^{(-)}(n_{\rm max}) = \Gamma_\chi^{(+)}(n_{\rm min})=0$ 
by hand that breaks the analytical $n$~-~dependence of equations~(\ref{System2}) and 
(\ref{Relation}) at the boundaries. The only exceptional situation, when the above way to 
solve the Eq.~(\ref{System2}) is indeed true, corresponds to the case $n_1 = n_{\rm min}$, 
$n_2 = n_{\rm max}$, with $n_{1(2)}$ being the zeros of functions $\Gamma^{(\pm)}_\chi(n)$.
In this case the substitution~(\ref{subst}) gives $p(n_{\rm max}+1) = p(n_{\rm min}-1) = 0$
and hence the actual values of $\Gamma^{(-)}_\chi(n_{\rm max}+1)$ and 
$\Gamma^{(+)}_\chi(n_{\rm min}-1)$ in Eq.~(\ref{System2}) play no role. Then we arrive to 
the action $S(\chi) = t_0 \Lambda(\chi)$ in the form which was claimed in section V~(a).


\end{section}


\begin{thebibliography}{99}

\bibitem{BlanterReview} 
Ya. M. Blanter and M. B\"{u}ttiker, Phys. Rep. {\bf 336}, 1 (2000).















\bibitem{Levitov} L. S. Levitov and G. B. Lesovik, JETP Lett. {\bf 58}, 230
(1993); L. S. Levitov, H.-W. Lee, and G. B. Lesovik,
Journal of Mathematical Physics, {\bf 37} (1996) 4845.

\bibitem{Yakovets} H.\ Lee, L.\ S.\ Levitov, A.\ Yu.\ Yakovets, Phys. Rev. B, 
{\bf 51}, 4079 (1995)

\bibitem{Blanter} Ya. M. Blanter, H. Schomerus, and C. W. J. Beenakker, 
Physica E {\bf 11}, 1 (2001)

\bibitem{Andreev} A.\ Andreev and A.\ Kamenev, Phys. Rev. Lett., {\bf 85}, 1294 (2000)

\bibitem{Levitov1} L.\ S.\ Levitov, arXiv: cond-mat/0103617

\bibitem{Mirlin} Y. Makhlin and A. D. Mirlin,  Phys. Rev. Lett., {\bf 87}, 276803 (2001) 

\bibitem{RefYuli} Yu. V. Nazarov, 
  Ann. Phys. (Leipzig) {\bf 8} Spec. Issue, SI-193 (1999), cond-mat/9908143.

\bibitem{General} Yu. V. Nazarov, {\it Generalized Ohm's Law}, in: 
Quantum Dynamics of Submicron Structures, eds. H. Cerdeira, B. Kramer, G. 
Schoen, Kluwer, 1995, p. 687.

\bibitem{Belzig} W.\ Belzig and Yu.\ V.\ Nazarov, Phys. Rev. Lett., {\bf 87}, 067006 (2001);
W.\ Belzig and Yu.\ V.\ Nazarov, Phys. Rev. Lett., {\bf 87}, 197006 (2001).

\bibitem{NazBag} Yu.\ V.\ Nazarov and D.\ A.\ Bagrets, Phys. Rev. Lett., {\bf 88}, 196801 
(2002)

\bibitem{Borlin} J.\ Borlin, W.\ Belzig,  and C.\ Bruder, Phys. Rev. Lett., 
{\bf 88}, 197001 (2002)

\bibitem{QBReview} 

\bibitem{Korotkov} A.\ N.\ Korotkov, Phys. Rev. B, {\bf 49}, 10 381 (1994)

\bibitem{Hershfield} S.\ Hershfield, J.\ H.\ Davis, P.\ Hyldgaard, C.\ J.\ Stanton,
and J.\ Wilkins, Phys. Rev. B, {\bf 47}, 1967 (1993)

\bibitem{Hanke} U.\ Hanke, Yu.\ M.\ Galperin, K.\ A.\ Chao, N.\ Zhou, Phys. Rev. B,
{\bf 48}, 17 209 (1993)

\bibitem{Galperin} U.\ Hanke, Yu.\ M.\ Galperin, K.\ A.\ Chao, Phys. Rev. B,
{\bf 50}, 1595 (1994)

\bibitem{Bulka} B.\ R.\ Bulka, J.\ Martinek, G.\ Michalek, J.\ Barna\'s,
Phys. Rev. B, {\bf 60}, 12 246 (1999)


\bibitem{Birk} H.\ Birk, M.\ J.\ M.\ de Jong, and C.\ Sch\"onenberger, Phys. Rev. Lett.,
{\bf 75}, 1610 (1995) 

\bibitem{LarMat}
A.\ I.\ Larkin and K.\ A.\ Matveev,   
Zh.\ \'Eksp.\ Teor.\ Fiz.\ {\bf 93}, 1030 (1987)
[Sov.\ Phys.\ JETP {\bf 66}, 580 (1987)]

\bibitem{GlazMat}
L.\ I.\ Glazman and K.\ A.\ Matveev,   
Pis'ma Zh.\ \'Eksp.\ Teor.\ Fiz.\ {\bf 48}, 403 (1988)
[JETP Lett. {\bf 48}, 445 (1988)]

\bibitem{Savchenko} V.\ V.\ Kuznetsov, A.\ K.\ Savchenko, D.\ R.\ Mace, E.\ H.\ Linfield
and D.\ A.\ Ritchie, Phys.\ Rev.\ B {\bf 56}, R15 533 (1997) 

\bibitem{Beenakker} C.\ W.\ J.\ Beenakker, Phys. Rev. B. {\bf 44}, 1646 (1991)

\bibitem{IngoldNaz} 
 G.-L.\ Ingold, Yu.\ V.\  Nazarov, in {\it Single Charge Tunneling}, 
 NATO ASI Series B: {\bf 294}, ed. H.\ Grabert, M.\ H.\ Devoret (NewYork, 1992) 


\bibitem{vanKampen} 
N.G. van Kampen, {\it Stochastic processes in physcics and chemistry}, Rev. and enl. eddition, 
{Elsevier Scinece Publishes B.V., North-Holland, 1992}

\bibitem{Struben} Yu.\ V.\ Nazarov, J.\ J.\ R.\ Struben, Phys. Rev. B, {\bf 53}, 15 466 (1996)

\end{thebibliography}
\end{document}